\documentclass{aa}  

\usepackage{multirow}
\usepackage{pdflscape}
\usepackage{afterpage}
\usepackage{graphicx}
\usepackage{txfonts}
\usepackage{hyperref}
\newcommand{\revision}[1]{#1}

\begin{document}

   \title{A Deep Neural Network Approach to Compact Source Removal}


   \author {M. Madarász\inst{\ref{konkoly},\ref{mtaexcellence}}
          \and
        G. Marton\inst{\ref{konkoly},\ref{mtaexcellence}}
          \and
          I. Gezer\inst{\ref{konkoly},\ref{mtaexcellence}}
          \and
          S. Lehner\inst{\ref{ellis},\ref{jku}}
          \and
          J. Roquette\inst{\ref{inst-ch-obs}}
          \and
          M. Audard\inst{\ref{inst-ch-obs}}
          \and
          D. Hernandez\inst{\ref{univie}}
          \and          
          O. Dionatos\inst{\ref{nhmvie}}
          }

   \institute{
   Konkoly Observatory, Research Centre for Astronomy and Earth Sciences, Hungarian Research Network (HUN-REN), H-1121 Budapest, Konkoly Thege Miklós út 15-17., Hungary\label{konkoly}\\
              \email{madarasz.mate@csfk.org}
\and
CSFK, MTA Centre of Excellence, Budapest, Konkoly Thege Miklós út 15-17., H-1121, Hungary\label{mtaexcellence}
\and ELLIS Unit Linz\label{ellis}
\and Institute for Machine Learning, Johannes Kepler University Linz, Austria\label{jku}
\and
Department of Astronomy, University of Geneva, Chemin Pegasi 51, 1290 Versoix, Switzerland\label{inst-ch-obs}
\and
Department of Astrophysics, University of Vienna, Türkenschanzstrasse 17, 1180 Vienna, Austria\label{univie}
\and Natural History Museum Vienna, Burgring 7, 1010 Vienna, Austria\label{nhmvie}
}

   \date{Received Day Month Year; accepted Day Month Year}

 
  \abstract
   {Analyzing extended emission in photometric observations of star-forming regions requires maps free from compact foreground, embedded, and background sources, which can interfere with various techniques used to characterize the interstellar medium. Within the framework of the NEMESIS project, we apply machine learning techniques to improve our understanding of the star formation timescales, which involves the unbiased analysis of the extended emission in these regions.}
   {We present a deep learning-based method for separating the signals of compact sources and extended emission in photometric observations made by the \textit{Herschel} Space Observatory, facilitating the analysis of extended emission and improving the photometry of compact sources.}
   {Central to our approach is a modified U-Net architecture with partial convolutional layers. This method enables effective source removal and background estimation across various flux densities, using a series of partial convolutional layers, batch normalization, and ReLU activation layers within blocks. Our training process utilized simulated sources injected into \textit{Herschel} images, with controlled flux densities against known backgrounds. A dynamic, signal-to-noise ratio-based adaptive masking system was implemented to assess how prominently a compact source stands out from the surrounding background.}
   {The results demonstrate that our method can significantly improve the photometric accuracy in the presence of highly fluctuating backgrounds. Moreover, the approach can preserve all characteristics of the images, including the noise properties. }
   {The presented approach allows users to analyze extended emission without the interference of disturbing point sources or perform more precise photometry of sources located in complex environments. We also provide a Python tool with tutorials and examples to help the community effectively utilize this method.}

   \keywords{ISM: structure - Techniques: image processing - Techniques: photometric - Methods: data analysis}

   \maketitle
%

\section{Introduction}


Stars form as the result of a long series of processes that take place within dense molecular clouds. The first step of the process is the gravitational collapse \citep{Shu:77,Shu:87a,Larson2003}. The resulting young stellar objects (YSOs) form young stellar groups or clusters, often organized in a “hub-filament structure”. \citet{Andre2014} proposed a paradigm suggesting that star formation occurs primarily within filaments, implying a universal process extending up to galactic scales. Further research has identified even smaller structures within filaments, known as “fibers” \citep{Hacar2013, Hacar2018, Hacar2024,Lee2014,Zavagno2023}, which exhibit similar dynamical properties across both low- and high-mass star-forming regions. 

The structure of the interstellar medium (ISM) shows a great variety across the Milky Way. It is shaped by many physical processes, including heating, cooling, gravity and magnetic fields. We already know that star formation efficiency and star formation rates also show significant differences in different environments. \citet{2012ApJ...761..156F} discussed that observations of star-forming clouds reveal that the star formation rate (SFR) surface density spans four orders of magnitude and positively correlates with the gas surface density \citep{2010ApJ...723.1019H}, indicating that denser gas supports a higher rate of star formation. This raises a key question: How is gas locally compressed in the ISM to form dense cores that eventually become gravitationally unstable, leading to star formation? Gas compression through shocks, driven by large-scale supersonic turbulence, may play a critical — if not the primary — role in establishing the initial conditions for star formation \citep[see, e.g., reviews by][]{2004RvMP...76..125M,2007ARA&A..45..565M}. This information can be inferred from the analysis of the extended emission.

To analyze extended emissions, maps need to be free from contaminating sources. Even in Fourier space, compact sources (those with extensions similar to or smaller than the beam’s FWHM at a specific wavelength) add substantial power to the image’s power spectrum across a broad range of spatial frequencies, affecting the image especially at frequencies near the beam size. Depending on the source density and clustering strength, lower spatial frequencies may also be contaminated, albeit with lower power density spread over a wide bandwidth. 

Image analysis techniques include power spectrum analysis that quantifies the spatial distribution of fluctuations in the ISM by measuring how intensity varies across different spatial scales. It is widely used to study turbulence in the ISM, with different scales indicating the presence of large molecular clouds, small clumps, or filamentary structures. Power spectra help identify whether structures follow a particular scaling law \citep[see, e.g.][]{2015A&A...584A.111R}.

The probability density functions (PDFs) represent the distribution of intensities (such as column densities or densities) in a region of the ISM, showing the frequency of different densities. PDFs are useful for distinguishing different physical regimes within the ISM, like a log-normal PDF often indicates turbulence-dominated gas, while power-law tails can suggest gravitational collapse or strong density contrasts, such as those in star-forming regions \citep[see, e.g.][]{2024MNRAS.532.4661K}.

$\Delta$-variance \citep{1998A&A...336..697S}, introduced as a wavelet-based measure for the statistical scaling of structures in astronomical maps, is a tool for quantifying spatial scaling and structure, particularly helpful in characterizing turbulence. It measures how variance changes with scale, often providing more sensitivity to large-scale structures than power spectrum analysis. $\Delta$-variance is particularly effective for analyzing hierarchical structures and fractal-like features in the ISM, helping to differentiate between turbulent and non-turbulent regions. One of its applications to the analysis of extended emission is presented by \citet{2014ApJ...788....3E}. 

These are the most common techniques, but several others are used in the different sub-fields of astrophysics, recently discussed and compared by \citet{2024arXiv240714068M}.
In any case, comparing image analysis techniques becomes challenging without source subtraction, as different methods vary in their sensitivity to discrete substructures. Techniques relying on sparsity information can be skewed by just a few point sources, while methods analyzing entire intensity maps are more affected by clustering. Effective analysis of extended emission requires a source subtraction technique that targets a specific range of spatial frequencies while preserving the noise properties of the image. There have been several approaches for source extraction in the presence of structured background and for the removal of sources. \citet{2021A&A...649A..89M} developed the getsf code, \citet{2011A&A...530A.133M} the CuTEx code, and \citet{2014ExA....37..347M} the boloSource() code, all of them using different approaches, but based on traditional mathematical and physical models.

In this paper, we present a deep-learning tool, designed to remove point sources from the photometric observations of the PACS instrument \citep{2010A&A...518L...2P} on-board of the \textit{Herschel} Space Observatory \citep{2010A&A...518L...1P}. This work is part of our Novel Evolutionary Model for the Early Stages of stars with Intelligent Systems (NEMESIS\footnote{https://nemesis.univie.ac.at}) project, and a continuation and extension of the new \textit{Herschel}/PACS catalog \citep{point_source_catalogue}. The tool can be trained for the photometric data of other instruments on other spacecrafts, as well, thanks to the modified U-Net \revision{\citep{UNet}} architecture with partial convolutional layers integrated within an inpainting framework, central to our approach. Section~\ref{section:data} outlines the data used across the different stages of preprocessing, training, and validation. A detailed description of the deep-learning algorithm, including its architecture, implementation, and the photometric methods employed for validation, is provided in Sect.~\ref{section:methods}. The performance of the method is then presented in Sect.~\ref{section:results}, followed by a discussion in Sect.~\ref{section:discussion} on photometric accuracy and the image properties of the resulting data after applying our approach to various examples. 

\section{Data}\label{section:data}

In this section, we provide a detailed explanation of the data used in this study and its key properties. We describe how we created simulations for the task, how the dataset was split into training, validation, and test sets, the preprocessing steps we applied, and how we constructed the masks for the compact sources.

This problem requires careful preparation to approach it as a supervised learning problem. The main idea is straightforward: to remove an object from an image, we need to understand what lies behind it. To create training data, we use original images where the true background is already known and sources are then injected into these backgrounds, simulating the presence of sources as seen in real observational data. This approach provides paired data: the input image, which includes the added sources on the background, and the ground truth target image, which is the original background without the sources. It allows us to construct a dataset where the machine learning model can learn to predict the clean background from the input image containing compact sources. In the subsections that follow, we explain these steps in detail.

\subsection{\textit{Herschel} Science Archive (HSA)}
\label{section:hsa}

The \textit{Herschel} Space Observatory \citep{2010A&A...518L...1P} was the fourth cornerstone mission in the European Space Agency (ESA) science program. More than 35,000 observations were made during its mission. The different data products created by the Standard Product Generator (SPG) pipelines are available through the \href{http://archives.esac.esa.int/hsa/whsa/}{\textit{Herschel} Science Archive} (HSA) and can be downloaded via the web interface or through the \href{https://www.cosmos.esa.int/web/herschel/hipe-download}{HIPE} software, specifically developed for \textit{Herschel} data manipulation. 

The PACS photometer was a dual-band instrument composed of two filled silicon bolometer arrays: the Blue camera, consisting of $32 \times 64$ pixels, and a Red Camera, consisting of $16 \times 32$ pixels. The Blue camera provided two filters centered at $70 \mu \mathrm{m}$ and $100 \mu \mathrm{m}$, (commonly referred to as BS - blue, and BL - green bands), while the Red camera acquired observations at the nominal wavelength of $160 \mu \mathrm{m}$ (R - red band). 


An important parameter of the PACS observations that we need to introduce is the Observing Mode, as we used two types of data. The Observing Mode could either be the nominal Scan Map or Parallel Mode. In the latter, the observations were acquired together with the SPIRE instrument with a simultaneous five-band photometry of the same field of view. In both modes, the bolometer read-out frequency was 40 Hz, but due to data-rate limitations, an on-board averaging of 4 frames was performed, except for the blue and green bands in Parallel Mode, where the averaging was increased to 8 frames. For our simulations, we used both Scan Map and Parallel Observing Modes for all the three bands, resulting in a total of six different types of data.

The standard satellite scanning speed was 20$^{\prime\prime}$s$^{-1}$. The highest speed of 60$^{\prime\prime}$s$^{-1}$ was dedicated to large Galactic surveys and Parallel Mode maps.

For our analysis, we exclusively utilized Level 2.5 and Level 3 maps created with the JScanam algorithm \citep{JScanam}. These maps are well-suited for our purposes, as the JScanam method ensures reliable recovery of both extended emission and point-like sources without relying on filtering or noise models.

\subsection{Point Spread Function}
\label{section:psf}

\begin{figure}[!htb]
   \centering
   \includegraphics[width=\hsize]{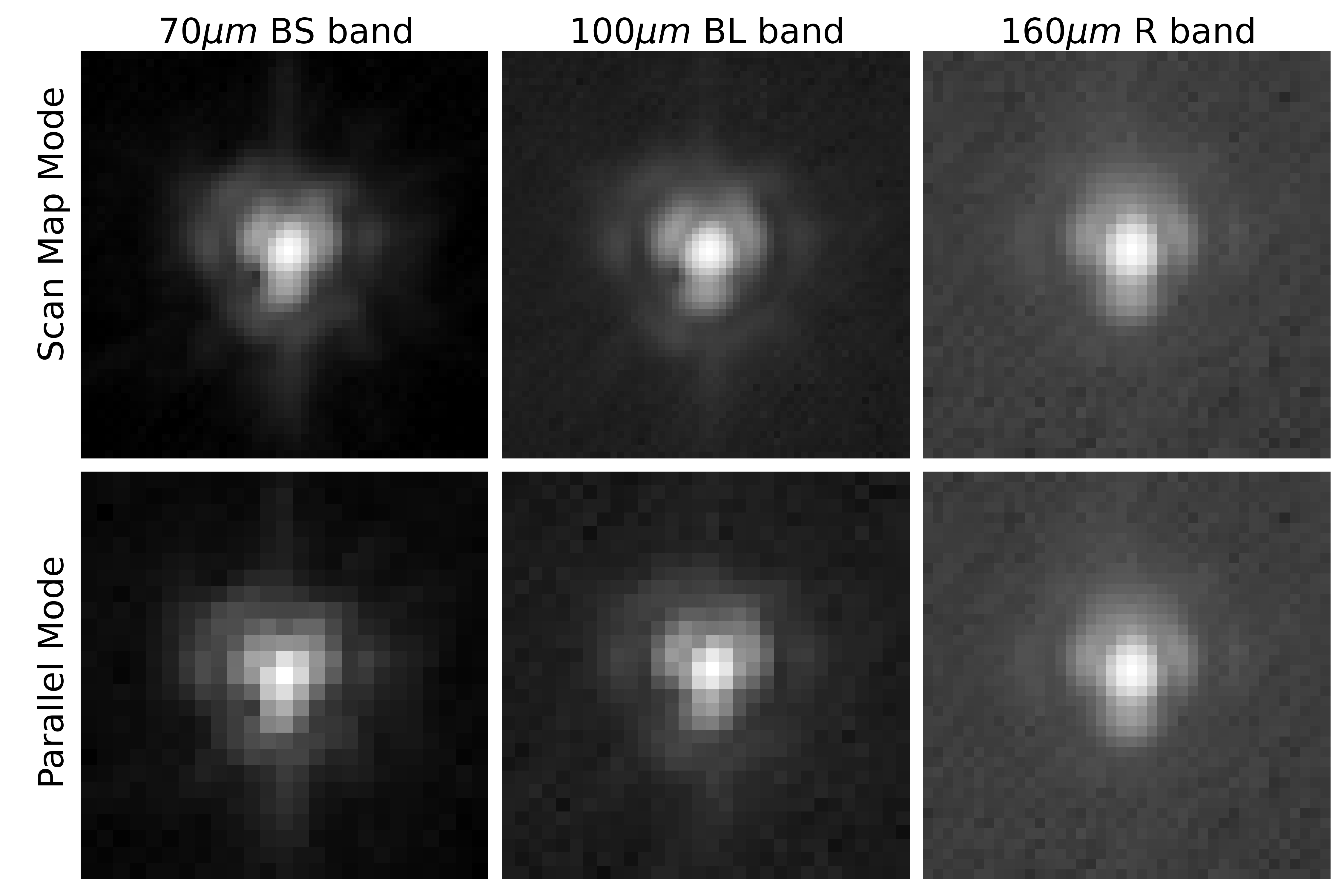}
    \caption{The figure shows the six Vesta PSF images after they have been transformed to match the resolution of the observation image data. The first row represents the Scan Map mode, the second is the Parallel mode, and the three columns correspond to the blue, green and red bands, respectively. Note that all the images were plotted to have the same size for visualization purposes, but it can be seen that the resolution of the PSF images differs between the bands and Observing Modes.}
    \label{fig:psf_comparison}
\end{figure}

The shape of the PACS Point Spread Function (PSF) is dominated by a tri-lobe pattern in all three bands. A detailed description can be found in the PACS photometer point spread function technical note \citep{PACS_TechNote}. For our simulations, to simulate real sources, we used the official Vesta PSFs, which are specific to the Observing Mode, band, and scan speed \citep{PACS_TechNote}. We transformed the PSF images to match the resolution of the observations, resulting in the six PSF images, as shown in Fig.~\ref{fig:psf_comparison}.

For 60$^{\prime\prime}$s$^{-1}$ scan speeds, the shape of the PSF is more elongated. However, with different scan angles, the elongation occurs in different directions, which can be handled using circular masks, as discussed in Sect.~\ref{section:mask}.

\subsection{Simulations}

As mentioned earlier, with the two Observing Modes and the three bands, we have a total of six different types of data to process. However, since the three bands within the modes are more similar in appearance in the image cutouts, we decided to merge the data for each mode. This way, we only need to train two models, one for the Scan Map mode and one for the Parallel mode, and allow the network to learn the differences between the bands using masks designed specifically for the bands, as described in Sect.~\ref{section:mask}. The need to handle Scan Map and Parallel modes separately arises from their resolutional differences.

\subsubsection{Data Generation Process}
\label{section:data_generation}

We covered a wide range of spectral flux densities to ensure that our model could handle a variety of extreme cases. Flux density levels, defined from $10$ mJy to $40,000$ mJy, were chosen using irregular steps. Denser steps were used at lower flux densities, as they typically exhibit greater variation against the background compared to higher values, which show less visual difference. The categories are shown in Table~\ref{table:flux_density_levels}. All the simulations were generated with the following steps:

\begin{enumerate}
    \item An image field was selected with relatively few real sources present.
    \item We determined how many random sources could be injected into the image to maximize the number of sources while avoiding large empty spaces and excessive clustering of sources in smaller areas. The choice of the number, ranging between 25 and 210, was loosely proportional to the image size, but manually adjusted for each map based on visual inspection.
    \item We defined the minimum and maximum flux density levels to be used, so that at the minimum level, the injected sources would start to become visible against the background, and at the maximum level, the sources would be realistically bright compared to the background. This varied based on the complexity and brightness of the background in the selected image field.
    \item We created as many simulation images for a specific field as there are flux density levels (defined in Table~\ref{table:flux_density_levels}) within the determined minimum and maximum values (inclusive). For instance, suppose we determine to use flux levels between 40 and 125 for an image field. If we then check the table for values within this range, we find 40, 60, 80, 100, and 125 (inclusive). Therefore, five simulations will be created.
    \item Since the PSF image arrays were normalized to have a flux density of exactly 1 Jy, we scaled it to match the desired flux density level by multiplying the values accordingly.
    \item We generated the specified number of random positions for the particular field and injected the scaled PSFs into these positions.
    \item A corresponding \texttt{.csv} file was also created, which included the equatorial coordinates of right ascension \revision{(R.A.)} and declination (Dec) in the J2000.0 reference frame of the injected sources, with the flux density levels used in mJy. 
\end{enumerate}

\begin{table*}[!hbt]
    \caption{The defined flux density levels from 10 mJy to 40,000 mJy that were used to create simulations.}
    \label{table:flux_density_levels}
    \centering
    \begin{tabular}{|c|c|c|c|c|c|c|c|c|c|c|c|c|}
    \hline
    \multicolumn{13}{|c|}{Flux Density (mJy)} \\
    \hline
    10   & 20   & 40   & 60   & 80   & 100  & 125  & 150  & 175  & 200  & 250  & 300  & 350 \\
    400  & 500  & 600  & 700  & 800  & 900  & 1000 & 1250 & 1500 & 1750 & 2000 & 2500 & 3000 \\
    3500 & 4000 & 4500 & 5000 & 6000 & 7000 & 8000 & 9000 & 10000 & 11000 & 12000 & 13000 & 14000 \\
    15000 & 16000 & 18000 & 20000 & 22500 & 25000 & 27500 & 30000 & 32500 & 35000 & 37500 & 40000 &  \\
    \hline
    \end{tabular}
\end{table*}

\begin{figure}
   \centering
   \includegraphics[width=\hsize]{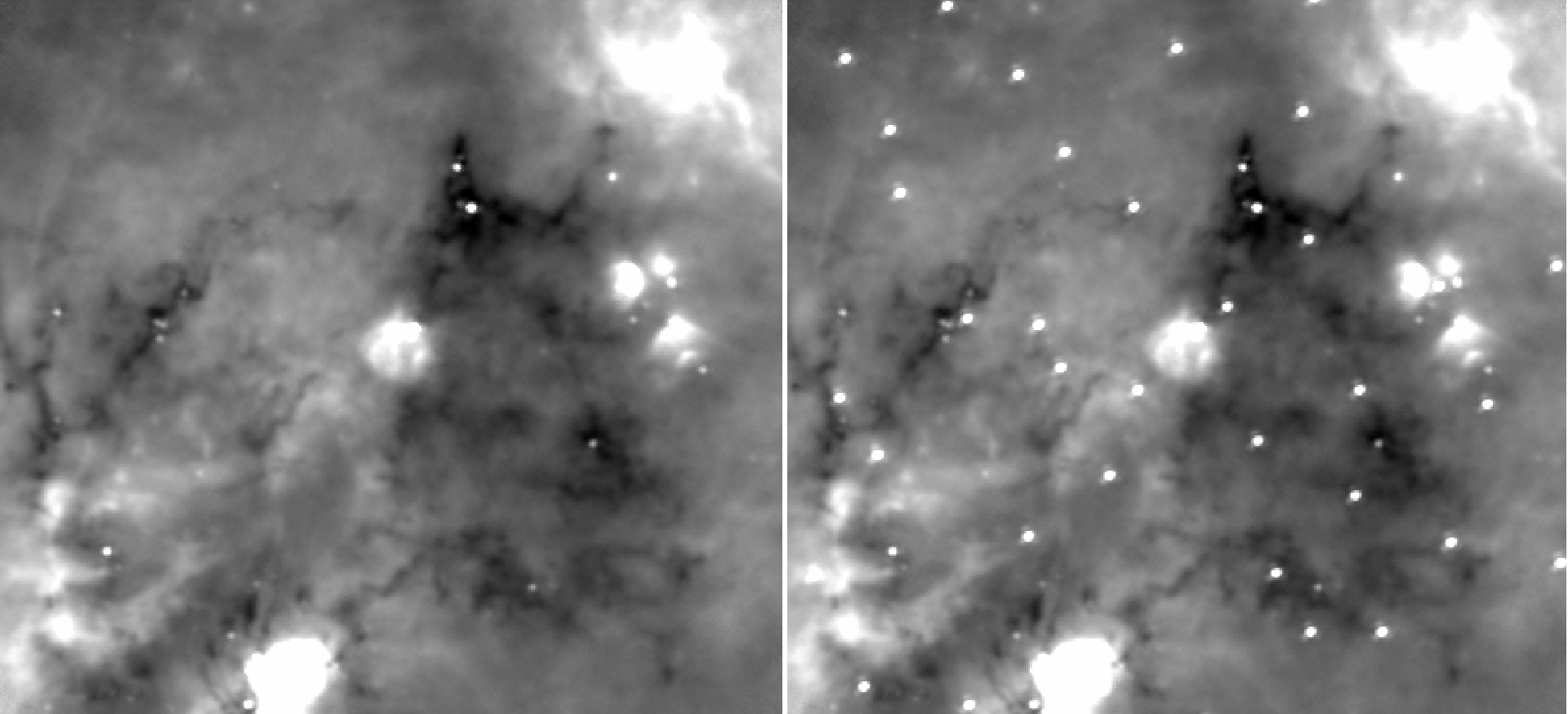}
    \caption{An example where a Scan Map Blue field with \texttt{obsid=1342191801} (on the left) was used as an input, and sources with a flux density of $2,000$ mJy were injected at random positions (on the right).}
    \label{fig:simulation_example}
\end{figure}

An example is shown in Fig.~\ref{fig:simulation_example}, where sources with flux density of 2,000 mJy were injected into the Scan Map Blue field with \texttt{obsid=1342191801}.

\subsubsection{Dataset Partitioning}
\label{section:train_test_split}

\begin{figure*}
	\resizebox{\hsize}{!}
	{\includegraphics[width=\hsize]{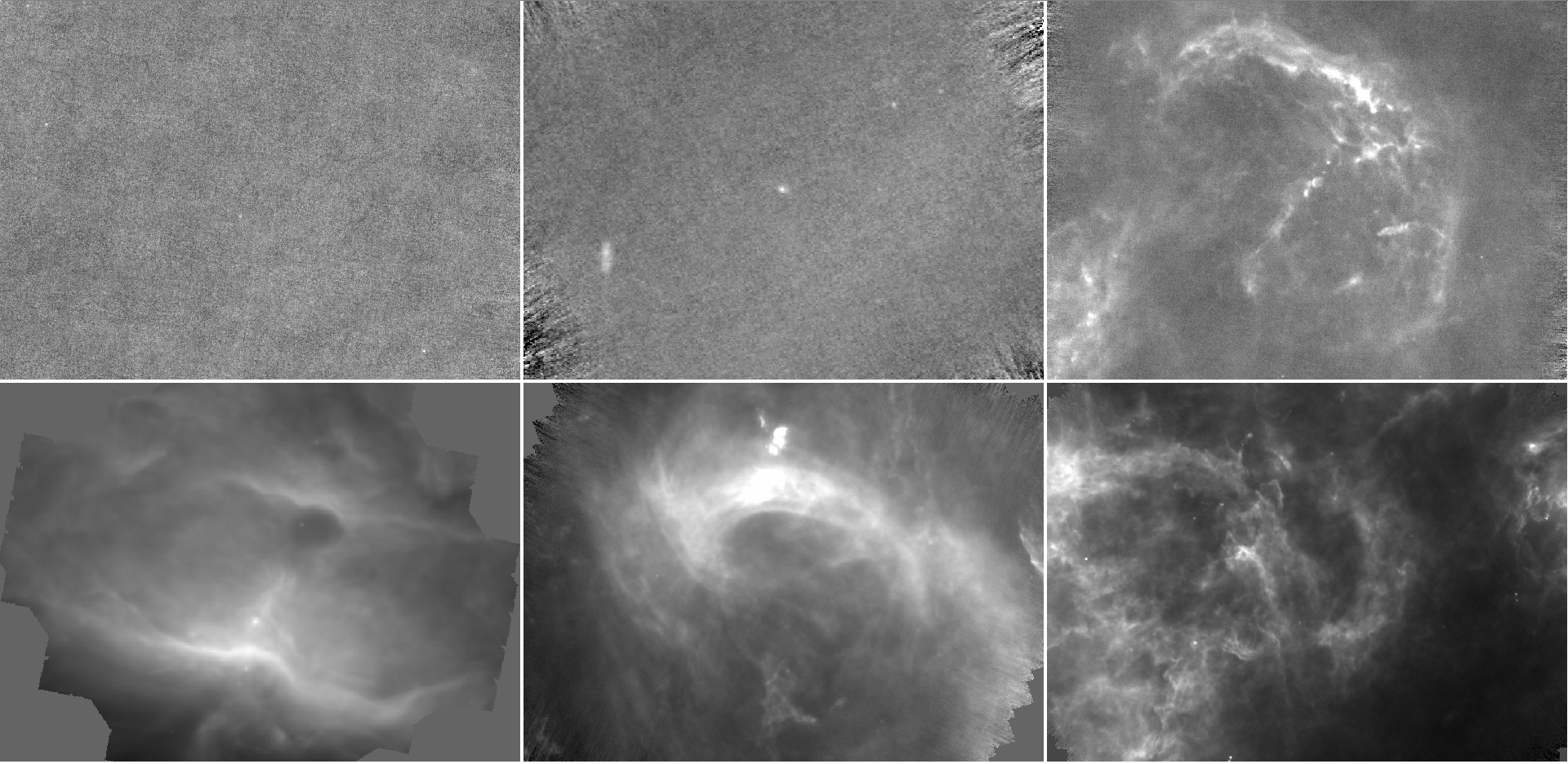}}
	\caption{The six fields used in the Scan Map Blue test dataset, categorized by approximate complexity. The upper left is the least complex, and complexity increases from left to right and top to bottom, with the bottom right being the most complex.}
	\label{fig:test_examples_complexity}
\end{figure*}

Several steps have been taken to create a balanced and well-representative dataset. First, we listed the observation identifiers (\texttt{obsid}s) of potential fields large enough to contain at least $25$ injected sources, and had only a few real sources present. We then sorted these \texttt{obsid}s into a few custom categories defined by us, based on a subjective assessment of their similarity in terms of complexity, appearance and structure. From these categories, we selected the most suitable fields. See the six examples in Fig.~\ref{fig:test_examples_complexity}, representing the six fields used in the Scan Map Blue test set. 

Next, we followed the steps described in Sect.~\ref{section:data_generation} to create the simulations for each selected \texttt{obsid} field. We then divided the list of \texttt{obsid}s into training, validation and test sets, ensuring that each of the three datasets contained at least one field from each of the custom categories. The training, validation and test sets are disjoint, since we used different \texttt{obsid}s in each set. Table~\ref{table:sources_and_fields} shows the distribution of sources and fields in the Scan Map and Parallel datasets, and additional details about the fields used can be found in Table~\ref{table:scanmap_blue_details}, Table~\ref{table:scanmap_green_details} and Table~\ref{table:scanmap_red_details} in the Appendix. The reason why we did not use simulations from the Parallel Red dataset for training is because it has the same PSF as the Scanmap version, thus only test simulations were created.

\begin{table*}[]
\caption{Total number of sources and fields for BS, BL, and R in the Scan Map and Parallel datasets.}
\label{table:sources_and_fields}
\centering
\begin{tabular}{c|cccccc||cccccc|}
\cline{2-13}
                                 & \multicolumn{6}{c||}{Scan Map}                                                                                                      & \multicolumn{6}{c|}{Parallel}                                                                                                                     \\ \cline{2-13} 
                                 & \multicolumn{3}{c|}{Number of Sources}                                            & \multicolumn{3}{c||}{Number of Fields}            & \multicolumn{3}{c|}{Number of Sources}                                            & \multicolumn{3}{c|}{Number of Fields}            \\ \cline{2-13} 
                                 & \multicolumn{1}{c|}{BS}     & \multicolumn{1}{c|}{BL}     & \multicolumn{1}{c|}{R}      & \multicolumn{1}{c|}{BS} & \multicolumn{1}{c|}{BL} & \multicolumn{1}{c||}{R}  & \multicolumn{1}{c|}{BS}     & \multicolumn{1}{c|}{BL}     & \multicolumn{1}{c|}{R}      & \multicolumn{1}{c|}{BS} & \multicolumn{1}{c|}{BL} & \multicolumn{1}{c|}{R}  \\ \hline
\multicolumn{1}{|c|}{Training}   & \multicolumn{1}{c|}{72,570} & \multicolumn{1}{c|}{72,345} & \multicolumn{1}{c|}{64,390} & \multicolumn{1}{c|}{29} & \multicolumn{1}{c|}{29} & \multicolumn{1}{c||}{23} & \multicolumn{1}{c|}{61,345}       & \multicolumn{1}{c|}{35,080}       & \multicolumn{1}{c|}{N/A}       & \multicolumn{1}{c|}{16}   & \multicolumn{1}{c|}{6}   & \multicolumn{1}{c|}{N/A}  \\
\multicolumn{1}{|c|}{Validation} & \multicolumn{1}{c|}{14,755} & \multicolumn{1}{c|}{17,095} & \multicolumn{1}{c|}{14,430} & \multicolumn{1}{c|}{7}  & \multicolumn{1}{c|}{7}  & \multicolumn{1}{c||}{4}  & \multicolumn{1}{c|}{15,230}       & \multicolumn{1}{c|}{9,485}       & \multicolumn{1}{c|}{N/A}       & \multicolumn{1}{c|}{4}   & \multicolumn{1}{c|}{3}   & \multicolumn{1}{c|}{N/A}  \\
\multicolumn{1}{|c|}{Test}       & \multicolumn{1}{c|}{18,095} & \multicolumn{1}{c|}{16,085} & \multicolumn{1}{c|}{15,490} & \multicolumn{1}{c|}{6}  & \multicolumn{1}{c|}{6}  & \multicolumn{1}{c||}{5}  & \multicolumn{1}{c|}{18,070}       & \multicolumn{1}{c|}{12,640}       & \multicolumn{1}{c|}{15,170}       & \multicolumn{1}{c|}{6}   & \multicolumn{1}{c|}{4}   & \multicolumn{1}{c|}{6}  \\ \hline
\end{tabular}
\end{table*}

\subsection{Preprocessing Workflow}
\label{section:preprocessing}

We wanted our network to work on specified input coordinates rather than the entire image field. Image fields vary significantly in spatial dimensions, number of sources, and other properties, making them difficult to use directly as inputs. Additionally, to ensure photometric accuracy, the central coordinates of the sources were essential. Also, to prevent the model from unintentionally modifying parts of the image other than the compact sources, it was essential to focus the network's operations locally around the specified coordinates. For these reasons, we used the given coordinates to extract smaller image cutouts centered on the compact sources. Below, we discuss the steps taken to prepare these inputs.

\subsubsection{Image Cutout Preparation}

For training, it is more efficient to prepare the image cutouts in advance and save them as \texttt{.npy} files, a binary format used by NumPy \citep{numpy} for storing arrays efficiently. This approach saves time during training, as generating the cutouts on-the-fly would be time consuming. Instead, we can simply load the pre-saved \texttt{.npy} files during training. However, in evaluation and production modes, the input consists of only the entire image maps and specified coordinates, as this is the simplest format for users to prepare. The image cutouts are then generated dynamically during the workflow.

We experimented with different cutout window sizes, starting with smaller sizes, such as $32 \times 32$ pixels, and then increasing them. Our aim was to find a size large enough to capture sufficient background information for the model to effectively extract features and remove even the brightest sources as well. We found that $96 \times 96$ pixel images offer a good balance between network complexity, and providing enough background context for the network to successfully remove the sources and estimate the underlying background.

\subsubsection{Normalization}

In our preprocessing workflow, we chose per-image normalization instead of global normalization. The pixel intensities in our images represent flux values, and the dynamic range of these fluxes can vary significantly between images due to the presence of objects with vastly different brightness levels.

Global normalization, which applies the same scaling factors across the entire dataset resulted in worse results and performance during training compared to the per-image version. Extremely bright regions dominate the normalization process, causing the average or dimmer point sources and background information to be poorly represented. Logarithmic scaling did not improve the performance, with the network's learning process remaining stagnant.

By using per-image normalization, each image is scaled individually based on its own minimum and maximum intensities. This allows the model to focus on the specific characteristics of each input image without being distorted by extreme flux variations present in the dataset. This approach also improves the network's ability to generalize across different fields with varying intensity profiles, ensuring better overall performance.

\subsubsection{Data Augmentation}

Since the input coordinates may not always be perfectly centered on the source, we applied random translations, with a maximum shift of 2 pixels in both vertical and horizontal directions, to mitigate this issue. Additionally, we applied horizontal and vertical flips, as well as random rotations, to augment the data and introduce sources with varying orientations. This augmentation strategy helps improve the model's robustness by introducing it to a wider variety of source positions and orientations. We implemented these transformations using PyTorch's \texttt{torchvision} library \citep{pytorch}.

\subsubsection{Determining Mask Size}
\label{section:mask}

For our setup, it was important to determine the appropriate shapes and sizes of the masks, which distinguish the regions of compact sources to be modified (removed) from the valid background.

At scan speeds of 60$^{\prime\prime}$s$^{-1}$, the PSF shape becomes more elongated. Due to varying scan angles, this elongation occurs in different directions, which can be addressed by employing circular masks. We experimented with single circular masks of different sizes, but this approach is not ideal. Bright sources, compared to their background, require larger masks, while faint objects, barely distinguishable from the background, need smaller masks. If a smaller mask is used for brighter sources, a significant portion of the PSF visible against the background will remain untreated and therefore not be removed. On the other hand, using large masks for faint sources covers more than necessary, reducing the valid region available for the network to work with.

\begin{figure}
\centering
\includegraphics[width=\hsize]{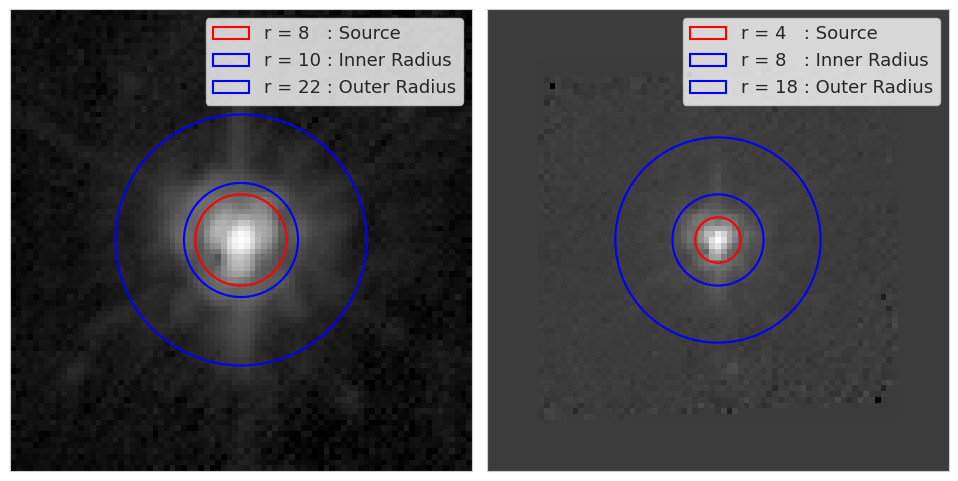} 
\caption{An illustration of the SNR calculation, with red and blue circles indicating the source region and the surrounding annulus, respectively. For the examples, we visualized the Scan Map Blue PSF on a logarithmic scale on the left, and the Parallel Blue PSF on a logarithmic scale on the right.} \label{fig:snr_calculation}
\end{figure}

To address this, we implemented dynamically calculated mask sizes for each source. To quantify how prominently a compact source stands out from the surrounding background, we calculated the signal-to-noise ratio (SNR) based on the peak flux of the source region. The peak intensity is effective for assessing contrast against the background, particularly for fainter sources. An illustration of the SNR calculation is shown in Fig.~\ref{fig:snr_calculation}. Given the different resolutions between the Scan Map and Parallel modes, the sizes of the source circle and surrounding annulus were adjusted accordingly. Specifically, for the Scan Map mode ($1.6 \times 1.6$ arcseconds per pixel), we used radii of 8 pixels for the source circle and 10 and 22 pixels for the inner and outer annulus, respectively. For the Parallel mode ($3.2 \times 3.2$ arcseconds per pixel), we used 4 pixels for the source circle and 8 and 18 pixels for the annulus. The exception is the Parallel Red mode, which shares the same PSF as the Scan Map mode, so we applied the Scan Map parameters for the SNR calculations in that case.

This annulus provides a good estimate of the background by computing both the mean $\left(\mu_{\text{bg}}\right)$ and standard deviation $\left(\sigma_{\text{bg}}\right)$ of the pixel intensities within it. The SNR is then calculated as:
\begin{equation}
    SNR = \frac{F_{\text{peak}} - \mu_{\text{bg}}}{\sigma_{\text{bg}}}.
\end{equation}

A higher SNR indicates that the source is more distinguishable from the background, meaning a larger portion of the PSF is visible and influences the surrounding region.

We calculated the SNR values for $2,000$ random sources for each type of data and analyzed the relationship between the SNR values and the appropriate mask size for each PSF separately. The results showed that using six distinct mask sizes, based on SNR thresholds, provides enough flexibility to handle the main differences between sources while achieving good performance. Instead of mapping the SNR values directly to mask sizes with interpolation, we decided to use discrete thresholds. This is because background conditions can vary significantly, making the mapping between SNR and the appropriate mask size imprecise. For example, two sources with the same flux density and similar background conditions may still have different SNR values if one of them lies near a brighter cloud or another bright object within the annulus region. These slight fluctuations in SNR make it more reliable to use thresholded categories rather than a direct mapping from SNR to mask size. This approach also helps us handle extreme cases more effectively.

\begin{figure}
\centering
\includegraphics[width=0.9\hsize]{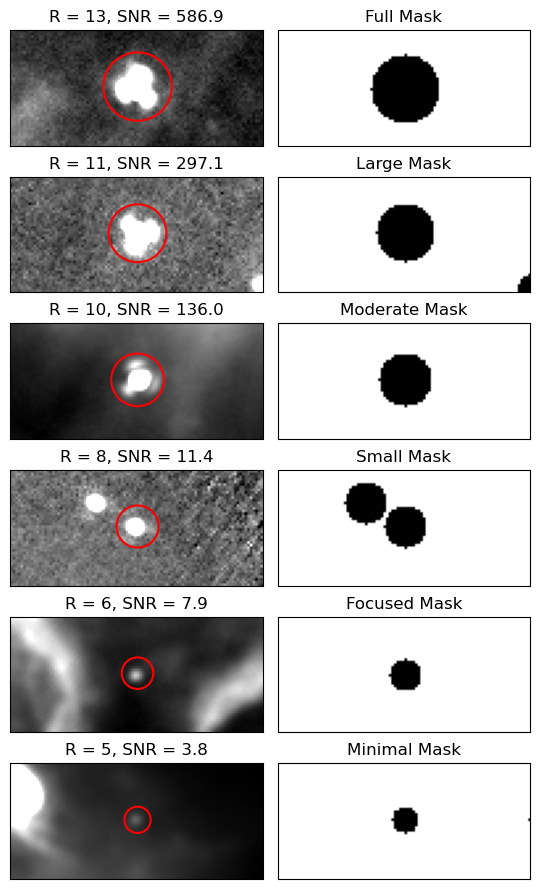}
\caption{Examples of the six different mask sizes used in the Scan Map Blue dataset. For each example, the left column shows the cutout image with the target object centered, along with the corresponding mask radius and the SNR value of the central object. The right column displays the binary mask used for inpainting. Note that multiple objects may appear within a cutout, but the SNR value and mask radius are determined based on the source located at the center of the image.}
\label{fig:mask_examples}
\end{figure}

We named the six masks, from largest to smallest, as: \texttt{Full Mask}, \texttt{Large Mask}, \texttt{Moderate Mask}, \texttt{Small Mask}, \texttt{Focused Mask}, and \texttt{Minimal Mask}. However, the appropriate mask sizes vary across different bands, thus the radii for the six masks were defined separately for each data type. The specific values for these masks are provided in Table~\ref{table:mask_sizes}. As an example, Fig.~\ref{fig:mask_examples} shows selected cases from the Scan Map Green dataset.

\begin{table*}[]
    \caption{Mask sizes (in pixel radius) corresponding to different SNR ranges for the datasets.}
    \label{table:mask_sizes}
    \centering
    \begin{tabular}{|ccccc|c|cccc|}
        \cline{1-5} \cline{7-10}
        \multicolumn{5}{|c|}{\textbf{Scan Map}}                                                                                                                                          &  & \multicolumn{4}{c|}{\textbf{Parallel}}                                                                                                      \\ \cline{1-5} \cline{7-10} 
        \multicolumn{1}{|c|}{\textbf{SNR Range}}     & \multicolumn{1}{c|}{\textbf{Mask Name}} & \multicolumn{1}{c|}{\textbf{Blue}} & \multicolumn{1}{c|}{\textbf{Green}} & \textbf{Red} &  & \multicolumn{1}{c|}{\textbf{SNR Range}}     & \multicolumn{1}{l|}{\textbf{Mask Name}} & \multicolumn{1}{c|}{\textbf{Blue}} & \textbf{Green} \\ \cline{1-5} \cline{7-10} 
        \multicolumn{1}{|c|}{SNR $\geq$ 300}         & \multicolumn{1}{c|}{Full Mask}          & \multicolumn{1}{c|}{12}            & \multicolumn{1}{c|}{13}             & 10           &  & \multicolumn{1}{c|}{SNR $\geq$ 855}         & \multicolumn{1}{c|}{Full Mask}          & \multicolumn{1}{c|}{9}             & 10             \\ \cline{1-5} \cline{7-10} 
        \multicolumn{1}{|c|}{140 $\leq$ SNR $<$ 300} & \multicolumn{1}{c|}{Large Mask}         & \multicolumn{1}{c|}{11}            & \multicolumn{1}{c|}{11}             & 9            &  & \multicolumn{1}{c|}{590 $\leq$ SNR $<$ 855} & \multicolumn{1}{c|}{Large Mask}         & \multicolumn{1}{c|}{8}             & 9              \\ \cline{1-5} \cline{7-10} 
        \multicolumn{1}{|c|}{13 $\leq$ SNR $<$ 140}  & \multicolumn{1}{c|}{Moderate Mask}      & \multicolumn{1}{c|}{9}             & \multicolumn{1}{c|}{10}             & 8            &  & \multicolumn{1}{c|}{300 $\leq$ SNR $<$ 590} & \multicolumn{1}{c|}{Moderate Mask}      & \multicolumn{1}{c|}{7}             & 8              \\ \cline{1-5} \cline{7-10} 
        \multicolumn{1}{|c|}{8 $\leq$ SNR $<$ 13}    & \multicolumn{1}{c|}{Small Mask}         & \multicolumn{1}{c|}{7}             & \multicolumn{1}{c|}{8}              & 7            &  & \multicolumn{1}{c|}{20 $\leq$ SNR $<$ 300}  & \multicolumn{1}{c|}{Small Mask}         & \multicolumn{1}{c|}{6}             & 7              \\ \cline{1-5} \cline{7-10} 
        \multicolumn{1}{|c|}{4 $\leq$ SNR $<$ 8}     & \multicolumn{1}{c|}{Focused Mask}       & \multicolumn{1}{c|}{6}             & \multicolumn{1}{c|}{6}              & 6            &  & \multicolumn{1}{c|}{13 $\leq$ SNR $<$ 20}   & \multicolumn{1}{c|}{Focused Mask}       & \multicolumn{1}{c|}{5}             & 5              \\ \cline{1-5} \cline{7-10} 
        \multicolumn{1}{|c|}{SNR $<$ 4}              & \multicolumn{1}{c|}{Minimal Mask}       & \multicolumn{1}{c|}{5}             & \multicolumn{1}{c|}{5}              & 4            &  & \multicolumn{1}{c|}{SNR $<$ 13}             & \multicolumn{1}{c|}{Minimal Mask}       & \multicolumn{1}{c|}{4}             & 4              \\ \cline{1-5} \cline{7-10} 
    \end{tabular}
    \tablefoot{The Parallel Red band uses the same mask sizes as the Scan Map Red band.}
\end{table*}

In these masks, white pixels indicate areas considered as valid background, while the black region represents the area to be modified. These masks, while not being overly strict, guide the model in differentiating between regions that are likely to be part of the compact source and those that constitute valid background.

\subsubsection{Input and Target Pairs}
\label{section:input_and_target}

The approach is straightforward: we start with the entire field containing the injected sources, the corresponding target maps — these are the original images used to inject the sources — and a binary mask with the same size as the other two maps, incorporating the circular masks. For each coordinate, we extract smaller cutouts from all three maps: the image with the injected source centered, the corresponding binary mask, and the original image. The input to the network consists of the cutout of the image with the injected source at its center, along with the corresponding binary mask, while the cutout from the original image serves as the target. These input-mask-target pairs are used to train the network to learn how to remove the PSF and accurately estimate the background behind the source.

\section{Methods}\label{section:methods}

We developed a neural network architecture grounded in an inpainting approach to address the challenge of compact source removal and extraction. Inpainting refers to a technique used to reconstruct missing or masked parts of an image based on the surrounding information. For our method, we use dynamically constructed masks with the SNR values, as described in Sect.~\ref{section:mask}.

\subsection{Architecture}

The U-Net architecture by \cite{UNet}, originally designed for biomedical image segmentation, forms the backbone of our model. This architecture is characterized by its symmetric structure, comprising an encoder for downsampling and decoder for upsampling, connected by skip connections. This design is particularly effective in preserving spatial information throughout the network, a crucial aspect for our task to access background regions directly.

Building upon the foundational concept of U-Net, we also drew inspiration from the work of \cite{pix2pix}, where they demonstrated the potential of such architecture in image-to-image translation tasks through their Conditional Generative Adversarial Network (cGAN). Their methodology highlighted the effectiveness of U-Net in learning complex, conditional mappings between input and output image domains.

We incorporated the \texttt{Convolution-BatchNorm-ReLU} module proposed by \cite{blocks} in their DCGAN network. This configuration, which includes a sequence of convolutional layers followed by batch normalization (\cite{batchnorm}) and Rectified Linear Unit (ReLU) activation (\cite{ReLU}), is known for its efficiency in deep convolutional networks. This is what \cite{pix2pix} used in their network as well, however, in contrast to the configuration in their generator network, we replaced the traditional convolutional layers with partial convolutional layers (PConv)\footnote{\url{https://github.com/naoto0804/pytorch-inpainting-with-partial-conv}: \\ - We used and adapted this implementation for partial convolution.}, used in \cite{pConv}. These layers are great at handling irregular (non-rectangle) holes in inpainting tasks, allowing the network to focus only on valid pixels in the convolutional process, and the mask is also updated layer-by-layer. 

We opted for strided convolutions instead of traditional pooling layers for both downsampling and upsampling. This choice is guided by the aim to enable the network to learn its own spatial transformations \citep{stridedConv}. Differing from \cite{pConv}, where nearest neighbor upsampling was used, we implemented transposed partial convolutional layers for detailed spatial reconstruction. These layers enable the network to autonomously learn upsampling, providing flexibility and accuracy in handling the complex spatial structures of astronomical data.

\subsection{Implementation Details}

\subsubsection{Network Architecture}

\begin{figure*}[h!]
\resizebox{\hsize}{!}
{\includegraphics[width=\hsize]{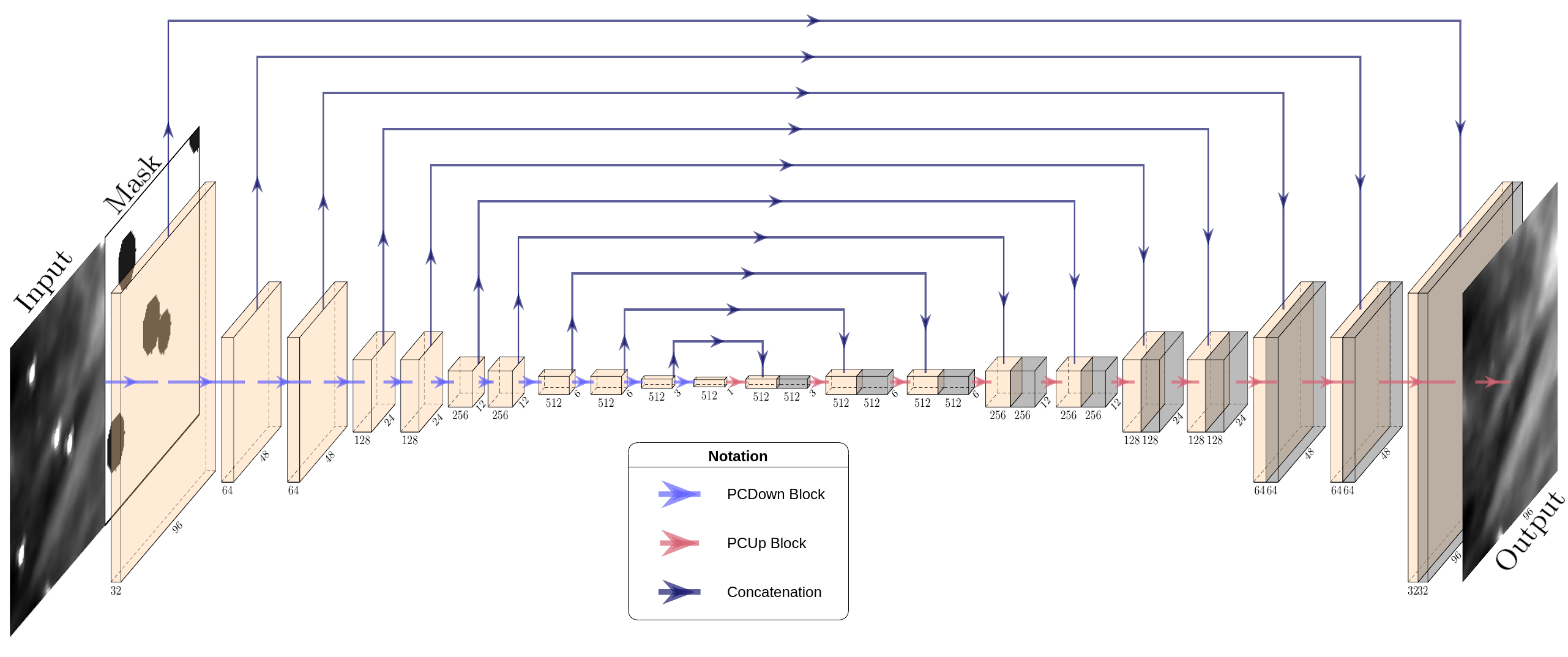}}
	\caption{Illustration of our architecture: This U-Net-like architecture comprises 11 PCDown Blocks in the encoder segment and an additional 11 PCUp Blocks in the decoder segment. The corresponding feature vectors at layer $i$ in the encoder are copied and concatenated to the input of layer $n-i$ in the decoder, where $n$ is the total number of layers. This structure helps the network in accessing specific regions of the input image directly. Further details of the architecture and its parameters can be found in Table~\ref{table:encoder_details}.}
	\label{fig:Architecture}
\end{figure*}

Our network, as depicted in Fig.~\ref{fig:Architecture}, consists of an encoder and a decoder segment. The network takes the input image along with the corresponding mask. This combined input then passes through the network, ultimately resulting in a one-channel grayscale image, where any compact source(s) present in the input are removed. A key element in both segments is the \texttt{Convolution-BatchNorm-ReLU} (CBR) block, which sequentially applies partial convolution, batch normalization, and an activation function. To construct our model, we defined two main types of these blocks, specifically designed for the encoder and the decoder.

\begin{figure}
	\centering
	\includegraphics[width=0.85\hsize]{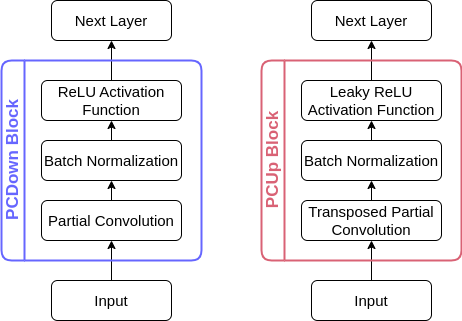}
	\caption{Core Components of the Architecture: The PCDown Block (left) downsamples and extracts features in the encoder, while the PCUp Block (right) upsamples in the decoder.}
	\label{fig:PCBlocks}
\end{figure}

In the encoder part, our Partial Convolution Downsample (PCDown) block configuration (left side in Fig.~\ref{fig:PCBlocks}) is constructed as follows:

\begin{itemize}
	\item \textbf{Partial Convolution} layer \citep{pConv} is used to handle the inpainting task, specifically to focus only on valid (background) regions of the input image during the convolution process. It dynamically updates the mask after each layer, allowing efficient feature extraction.
	\item \textbf{Batch Normalization} \citep{batchnorm} is applied after convolution, which stabilizes the learning process by normalizing the output of the previous layer.
	\item \textbf{ReLU Activation Function} \citep{ReLU} is employed to introduce non-linearity into the network, enabling it to learn more complex patterns in the data.
\end{itemize}

The Partial Convolution Upsample (PCUp) block (right side in Fig.~\ref{fig:PCBlocks}) in the decoder segment is similar but with some differences:

\begin{itemize}
	\item \textbf{Transposed Partial Convolution} serves the purpose of upsampling the feature maps. It combines the idea of transposed convolution \citep{tconv} with partial convolution.
	\item \textbf{Batch Normalization and Activation Functions} are similar to those in the encoder, however, we used LeakyReLU \citep{leaky1, leaky2} instead of ReLU with a negative slope parameter of 0.2.
\end{itemize}

\subsubsection{Encoder}

The encoder segment utilizes an alternating sequence of two block types to progressively downsample the input while extracting features:

\begin{itemize}
	\item \textbf{Strided Convolution Blocks} employ a PCDown block with a kernel size of 4, a stride of 2, and padding of 1. This configuration effectively reduces the spatial dimensions of the input image by half, enabling efficient downsampling by increasing the receptive field.
	\item \textbf{Feature Extraction Blocks} follow each strided convolution block. These employ a PCDown block with a kernel size of 3, a stride of 1, and padding of 1, maintaining the spatial dimensions while further extracting detailed features from the downsampled inputs.
\end{itemize}

The channel dimensions in these layers start from the number of input channels (1 in our case) and increase progressively, reaching up to 512. The last layer of the encoder, \texttt{PCDown11}, differs from the usual strided convolution blocks in order to reduce the spatial dimensions to $1 \times 1$ for $96 \times 96$ pixel input images, forming a compact encoded feature vector that captures the essential characteristics of the input. See the details of the encoder architecture in Table~\ref{table:encoder_details}.

\begin{table*}
    \caption{Details of Encoder and Decoder Architecture for 96x96 pixel inputs.}
    \label{table:encoder_details}
    \centering
    \begin{tabular}{|c|c|c|c|c|c||c|c|c|c|c|c|}
        \hline
        \multicolumn{6}{|c||}{Encoder Layers} & \multicolumn{6}{c|}{Decoder Layers} \\
        \hline
        Layer & In Ch. & Out Ch. & Kernel & Stride & Padding & Layer & In Ch. & Out Ch. & Kernel & Stride & Padding \\
        \hline
        PCDown1  & 1   & 32  & 3 & 1 & 1 & PCUp1 & 512  & 512 & 3 & 1 & 0 \\
        PCDown2  & 32  & 64  & 4 & 2 & 1 & PCUp2 & 1024 & 512 & 4 & 2 & 1 \\
        PCDown3  & 64  & 64  & 3 & 1 & 1 & PCUp3 & 1024 & 512 & 3 & 1 & 1 \\
        PCDown4  & 64  & 128 & 4 & 2 & 1 & PCUp4 & 1024 & 256 & 4 & 2 & 1 \\
        PCDown5  & 128 & 128 & 3 & 1 & 1 & PCUp5 & 512  & 256 & 3 & 1 & 1 \\
        PCDown6  & 128 & 256 & 4 & 2 & 1 & PCUp6 & 512  & 128 & 4 & 2 & 1 \\
        PCDown7  & 256 & 256 & 3 & 1 & 1 & PCUp7 & 256  & 128 & 3 & 1 & 1 \\
        PCDown8  & 256 & 512 & 4 & 2 & 1 & PCUp8 & 256  & 64  & 4 & 2 & 1 \\
        PCDown9  & 512 & 512 & 3 & 1 & 1 & \revision{PCUp9} & 128  & 64  & 3 & 1 & 1 \\
        PCDown10 & 512 & 512 & 4 & 2 & 1 & \revision{PCUp10} & 128  & 32  & 4 & 2 & 1 \\
        PCDown11 & 512 & 512 & 3 & 1 & 0 & \revision{PCUp11} & 64   & 1   & 3 & 1 & 1 \\
        \hline
    \end{tabular}
    \tablefoot{The encoder consists of 11 PCDown blocks, each employing a ReLU activation function. The decoder is composed of 11 PCUp blocks, using leaky ReLU activation function for all but the final block.}
\end{table*}

\subsubsection{Decoder}
\label{section:decoder}

The decoder segment of our network is designed to mirror the encoder's structure in terms of the spatial dimensionality of the feature maps (see Fig.~\ref{fig:Architecture}). This symmetry is crucial for the integration of skip connections, which connect each layer $i$ in the encoder and layer $n-i$ in the decoder, where $n$ is the total number of layers. Skip connections play an important role in preserving input details \citep{UNet}. In the decoder (see Table~\ref{table:encoder_details}), each layer is designed to match the parameters of the corresponding encoding layer. It comprises an alternating sequence of "upsampling blocks" and "feature refinement blocks":

\begin{itemize}
	\item \textbf{Upsampling Blocks} utilize PCUp blocks, where the transposed partial convolution is configured with a kernel size of 4, a stride of 2, and a padding of 1. This setup incrementally increases the spatial dimensions of the feature maps, effectively doubling them at each step.
	\item \textbf{Feature Refinement Blocks} follow each upsampling block, a PCUp block with a kernel size of 3, a stride of 1, and padding of 1 is employed. These blocks do not change the spatial dimensions but instead focus on refining and enriching the features that have been upsampled.
\end{itemize}

Additionally, as shown in the table, the number of input channels in the decoder is doubled due to the skip connections, where the output of the corresponding encoder layer is combined with the input of the matching decoder layer. This allows the decoder to have direct access to spatial information from earlier stages of the network, improving feature recovery and spatial consistency. The first layer of the decoder, \texttt{PCUp1}, diverges from the standard "Upsampling" and "Feature Refinement" block design to mirror the last block of the encoder, where we employed different parameters to compress the $96 \times 96$ pixel input into a $1 \times 1$ spatial dimension feature vector.

\subsection{Loss}

The loss function used during training plays a crucial role in guiding the model to produce accurate source-free images. Our loss function is an adaptation of the one proposed in \cite{pConv}, with some modifications customized to our specific requirements. We added the loss terms one-by-one and closely examined how each affected learning on a smaller subset of our dataset. In the following subsections, we describe the final loss terms used after our experiments.

\subsubsection{L1 Loss for Hole and Valid Regions}

Following the approach in \cite{pConv}, we compute the L1 loss separately for the hole (masked) and valid (unmasked) regions of the image. The L1 losses for masked and unmasked regions are defined mathematically as follows:

\begin{equation}
	\begin{split}
		\mathcal{L}_{\text{hole}} &= \frac{1}{N_{\mathbf{I}_\mathrm{gt}}} \left| \left| \left(1-\mathbf{M}\right) \odot \left(\mathbf{I}_{\text{out}} - \mathbf{I}_{\text{gt}}\right) \right| \right|_1 \\
		\mathcal{L}_{\text{valid}} &= \frac{1}{N_{\mathbf{I}_\mathrm{gt}}} \left| \left| \mathbf{M} \odot \left(\mathbf{I}_{\text{out}} - \mathbf{I}_{\text{gt}}\right) \right| \right|_1,
	\end{split}
\end{equation}

where $\mathbf{I}_\mathrm{out}$ is the output image generated by the network (the estimated background image), $\mathbf{I}_\mathrm{gt}$ is the ground truth image (the original background), and $\mathbf{M}$ represents the mask, with element-wise multiplication denoted by $\odot$. $N_{\mathbf{I}_\mathrm{gt}}$ denotes the total number of pixel values in an image, calculated as $C \times H \times W$, where $C$ is the channel size, and $H$ and $W$ are the height and width of the image $\mathbf{I}_{\text{gt}}$, respectively. $\mathcal{L}_{\text{hole}}$ targets the hole regions, where the sources are located, while $\mathcal{L}_{\text{valid}}$ ensures the integrity of the existing background is preserved.

\subsubsection{Style Loss}

The style loss, introduced by \cite{style_loss}, and utilized in the context of inpainting by \cite{pConv}, is also an integral part of our loss function. It helps in capturing the texture and style of the image, ensuring that the synthesized regions (background behind the source) blend seamlessly with the surrounding (valid) areas. The style loss is formulated as follows:

\begin{equation}
	\begin{split}
		\mathcal{L}_{\text{style}_\text{out}} &= \sum_{p=0}^{P-1} \frac{1}{C_p C_p} \left| \left| K_p \left( \left( \Psi_p^{\mathbf{I}_{\text{out}}} \right)^\top \left( \Psi_p^{\mathbf{I}_{\text{out}}} \right) - \left( \Psi_p^{\mathbf{I}_{\text{gt}}} \right)^\top \left( \Psi_p^{\mathbf{I}_{\text{gt}}} \right) \right) \right| \right|_1 \\
		\mathcal{L}_{\text{style}_\text{comp}} &= \sum_{p=0}^{P-1} \frac{1}{C_p C_p} \left| \left| K_p \left( \left( \Psi_p^{\mathbf{I}_{\text{comp}}} \right)^\top \left( \Psi_p^{\mathbf{I}_{\text{comp}}} \right) - \left( \Psi_p^{\mathbf{I}_{\text{gt}}} \right)^\top \left( \Psi_p^{\mathbf{I}_{\text{gt}}} \right) \right) \right| \right|_1
	\end{split}
\end{equation}

The Gram matrix, $\left( \Psi_p^{\mathbf{I}} \right)^\top \left( \Psi_p^{\mathbf{I}} \right)$, captures the correlation between different filter responses, effectively representing the style of the astronomical image by comparing their feature representations as extracted by a pre-trained convolutional neural network, like VGG-16 \citep{VGG-16}, where $\Psi_p$ is the feature map at layer $p$. The loss is calculated separately for the output image and a composite image ($\mathbf{I}_{\text{comp}}$), which is the raw output image, but with the non-hole pixels directly set to ground truth. The style loss then measures the difference in these correlations between the generated, source-free image ($\mathbf{I}_{\text{out}}$ or $\mathbf{I}_{\text{comp}}$) and the ground truth ($\mathbf{I}_{\text{gt}}$). $C_p$ is the number of channels in the feature map at layer $p$, and $K_p$ is a normalization factor.

\subsubsection{Structural Similarity Index Measure (SSIM) Loss}

We incorporate the Structural Similarity Index Measure (SSIM) loss (\cite{SSIM}) in our loss function, replacing the perceptual and total variation losses used in \cite{pConv}, as they did not improve background estimation and blending in our setting. The SSIM loss is advantageous in preserving the structural information in images, which is important when maintaining the integrity of dust cloud structures in such images. The SSIM loss is defined as:

\begin{equation}
	\mathcal{L}_{\text{SSIM}} = 1 - \text{SSIM}(\mathbf{I}_{\text{gt}}, \mathbf{I}_{\text{pred}}),
\end{equation}

where SSIM is the structural similarity index measure between the ground truth and the predicted image, and is defined as:

\begin{equation}
    \text{SSIM}(x,y) = \frac{\left( 2 \mu_x \mu_y + c_1 \right) \left( 2 \sigma_{xy} + c_2 \right)}{\left( \mu_x^2 + \mu_y^2 + c_1 \right) \left( \sigma_x^2 + \sigma_y^2 + c_2 \right)},
\end{equation}

where $\mu_x$ and $\mu_y$ are the pixel sample means, $\sigma_x^2$ and $\sigma_y^2$ are the variances of the images, $\sigma_{xy}$ is the covariance of the two image, $c_1 = \left( k_1 L \right)^2$ and $c_2 = \left( k_2 L \right)^2$ are two variables to stabilize the division with weak denominator and $L$ is the dynamic range of the pixel-values; $k_1 = 0.01$ and $k_2 = 0.03$ by default. We used the implementation from the Kornia library \citep{kornia}, a differentiable computer vision library for PyTorch that provides a wide range of tools seamlessly integrated into deep learning models.

\subsubsection{Source Loss}

\begin{figure}
	\centering
	\includegraphics[width=\hsize]{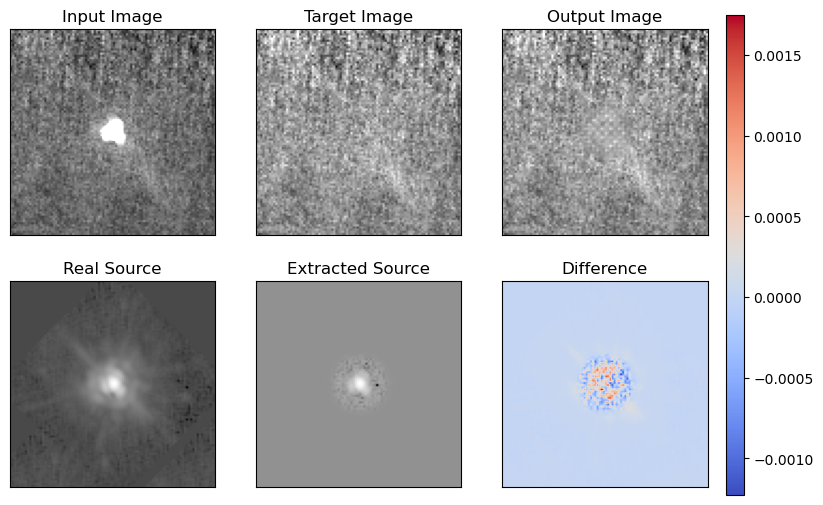}
	\caption{Visualization of the source loss. The top row displays the input image, the ground truth target (source-less image), and the output prediction (during the training process), respectively. The bottom row illustrates the real source (difference between the original image and the ground truth sourceless image), extracted source (difference between the original image and the predicted source-free image), and the difference between these two sources.}
	\label{fig:source}
\end{figure}

In addition to the previously described components, we included another loss to directly compare the original source and the extracted source, which we call the "source loss". This comparison ensures accurate extraction of the flux density and reconstruction of the PSF. The source loss can be expressed mathematically as follows:

\begin{equation}
	\mathcal{L}_{\text{source}} = \left| \left| \underbrace{\left( \mathbf{I}_\text{orig} - \mathbf{I}_\text{gt} \right)}_\text{real source} - \underbrace{\left( \mathbf{I}_\text{orig} - \mathbf{I}_\text{pred} \right)}_\text{extracted source} \right| \right|_1,
\end{equation}

where $\mathbf{I}_\text{orig}$ represents the original input image, $\mathbf{I}_\text{gt}$ is the ground truth sourceless image, and $\mathbf{I}_\text{pred}$ is the predicted source-free image. The difference between $\mathbf{I}_\text{orig}$ and $\mathbf{I}_\text{gt}$ reveals the real injected PSF (with the scaled flux values and rotated orientation), while the difference between $\mathbf{I}_\text{orig}$ and $\mathbf{I}_\text{pred}$ represents the source as extracted by the network, see the example in Fig.~\ref{fig:source}.

\subsubsection{Final Loss Function}

The final loss function for our network is a weighted sum of the individual loss components. We initially used the weights determined by \cite{pConv} for the components we also utilize, as they worked well. For the new components, we set their weights so that their values were of a similar magnitude to the others, ensuring they had a comparable impact. These values were calculated over a few epochs using different batches. For our final model, we used the following weighting:

\begin{equation}
	\mathcal{L}_{\text{total}} = 2\mathcal{L}_{\text{valid}} + 6 \mathcal{L}_{\text{hole}} + 120(\mathcal{L}_{\text{style}_{\text{out}}} + \mathcal{L}_{\text{style}_{\text{comp}}}) + 0.4 \mathcal{L}_{\text{SSIM}} + \mathcal{L}_\text{source}.
	\label{eq:final_loss}
\end{equation}

\subsection{Astropy and Aperture Photometry}
\label{section:aperture_photometry}

Astropy is a widely used open-source Python library for astronomy that provides a comprehensive set of tools for data manipulation, analysis, and visualization \citep{astropy:2013, astropy:2018, astropy:2022}.

In this study, we used Astropy’s \texttt{aperture\_photometry} function \citep{larry_bradley_2024_12585239}, which is designed to measure flux densities in defined regions of images. The \texttt{aperture\_photometry} function allows us to specify a circular aperture around a source, within which it sums the pixel values to estimate the source’s flux.

We also performed aperture photometry correction to account for the portion of the PSF that lies outside the aperture. This correction is important for obtaining a more accurate flux measurement, as parts of the PSF of the sources fall outside the defined aperture radius. \revision{Additionally, the shape of the PSF depends on the source's spectral energy distribution (SED). In our simulations, we used the PSF based on Vesta, which has an effective temperature that differs significantly from that of typical standard stars. To address this, we applied aperture corrections specific to Vesta for our simulated sources, while using different aperture corrections for comparison to standard star models in Sect.~\ref{section:std_stars_comparison} (see \citet{PACS_TechNote}).}

We applied corrected aperture photometry in two distinct ways. First, on the original image, we used a circular annulus around each source to estimate the mean background level, which we subtracted from the aperture photometry result to obtain the source flux. Then, to assess our model's performance, we measured the flux on the image where the source was extracted by our network. For this comparison, we performed aperture photometry with correction but without subtracting a background value, as our model estimates the background directly by removing the source.

\subsection{Training Procedure}

\subsubsection{Setup and Hyperparameters}

The training inputs were 1-channel images of size \(96 \times 96\) pixels, as detailed in Sect.~\ref{section:input_and_target}. The model was trained using a batch size of 64.

For the optimization process, we employed the Adam optimizer (\cite{adam}), known for its efficiency in handling sparse gradients and adaptive learning rates. The initial learning rate was set at 0.005. To enhance the training process and adapt the learning rate based on validation loss performance, we utilized the \texttt{ReduceLROnPlateau} scheduler (\cite{pytorch}) with a patience of 5 epochs and a reduction factor of 0.5. This approach helps in fine-tuning the learning rate, ensuring the model converges to an optimal solution efficiently.

\subsubsection{Epochs and Validation}

The training of our model followed a systematic approach, after each epoch of training on the training dataset, a validation loop was conducted on the separate validation dataset. During these validation loops, we closely observed the performance of the model using the final loss function Eq.~\ref{eq:final_loss}.

The learning rate adjustments were managed by the \texttt{ReduceLROnPlateau} scheduler, which was triggered based on the final loss observed on the validation dataset. This ensured that learning rate modifications were aligned with the overall performance improvements of the model.

To further refine the model, we implemented a checkpointing system where the model's parameters were saved each time a new lowest value of the final loss was achieved on the validation dataset, mitigating issues related to overfitting or training stagnation.



\section{Results}
\label{section:results}

Our main result is the open-source tool we developed and made available for the community to use in removing compact sources, hosted at our \href{https://github.com/xMediMadix/compact-source-removal}{GitHub repository} \footnote{\url{https://github.com/xMediMadix/compact-source-removal}}. Users can employ the pre-trained model for source removal from \textit{Herschel} images, or re-train the model on custom datasets. The repository includes three main configuration files, each designed for one of the tasks described below, along with a detailed \texttt{README.md} file explaining how to utilize the code effectively. In Sect.~\ref{section:discussion}, we will evaluate our model's performance across various aspects. This includes an accuracy comparison of the photometry pipeline we used on standard star models, a comparison between the measured flux densities of extracted sources and the known values of the injected PSFs, and an analysis of how the model impacts the power spectrum and the pixel intensity distributions of the maps.

\subsection{Removal on Simulation}

We provide resources to demonstrate our model’s evaluation on simulations, including examples to guide users through the process. The tool iterates through simulation files, applies source removal using the trained model, and exports the resulting source-free background images, along with \texttt{.csv} files that contain photometric results for each source at specified aperture sizes.

\begin{figure}[h!]
	\centering
	\includegraphics[width=\hsize]{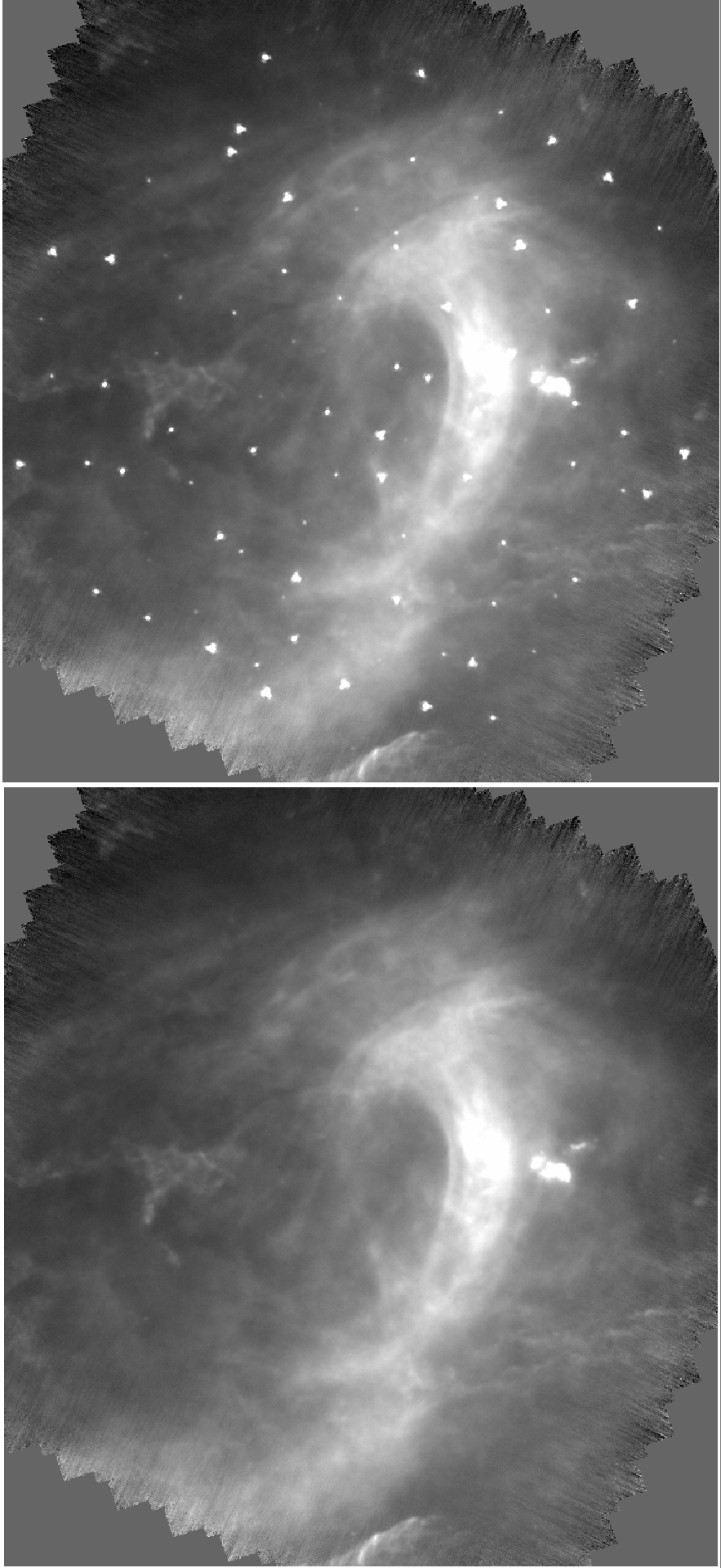}
	\caption{Example of simulated source removal. Top: Simulated image with 90 injected sources (randomly chosen flux densities from 40 to 10,000 mJy) within the Scan Map Blue observation data (\texttt{oid=1342237154}). Bottom: Model output with sources removed, providing a source-free background image.}
	\label{fig:sim_result_example}
\end{figure}

Figure~\ref{fig:sim_result_example} illustrates an example of this process. In this simulation, 90 sources with flux densities randomly chosen between 40 and 10,000 mJy were injected at random positions within the Scan Map Blue observation data with \texttt{oid=1342237154}. The top panel shows the simulated image with the injected sources, while the bottom panel displays the output of our evaluation tool, with compact sources removed to reveal the extended background emission.

\subsection{Removal on Real Data}

This core functionality is likely the most beneficial for the community. Users can specify parameters directly in a configuration file, defining aspects such as Observing Mode, band, aperture sizes for photometry, and other settings. We include examples to assist users. The tool processes input images by iterating through the coordinates of compact sources, removing them, and exporting a map with the sources removed, along with a \texttt{.csv} file that reports photometric results for the specified apertures. Users can input their own coordinates or use the new \textit{Herschel}/PACS Point Source Catalog created by \citet{point_source_catalogue}.

\begin{figure*}[!h]
	\resizebox{\hsize}{!}
	{\includegraphics[width=\hsize]{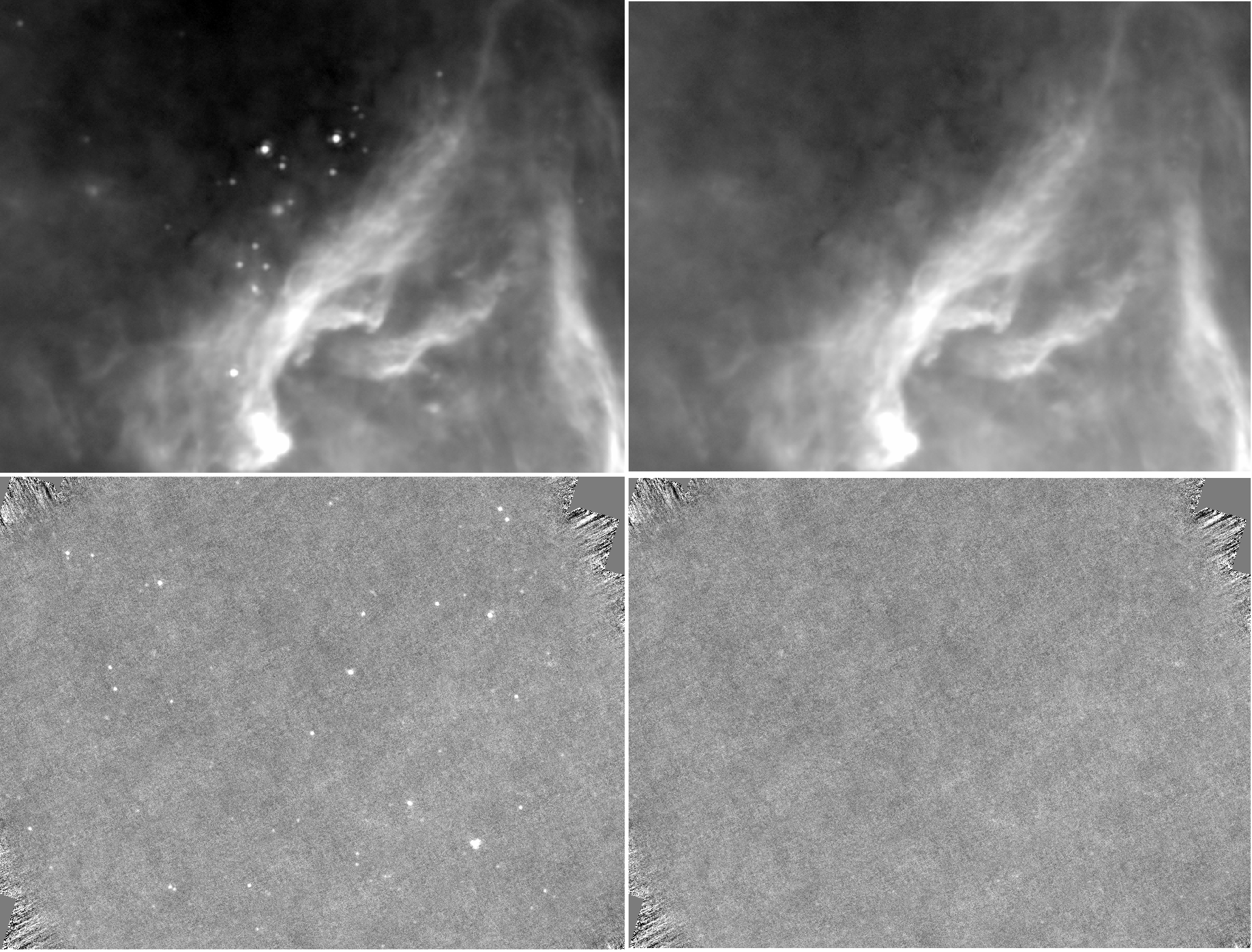}}
	\caption{Examples of compact source removal on real data. Top row: Original (left) and processed (right) Scan Map Green image (\texttt{oid=1342185553}). Bottom row: Original (left) and processed (right) Scan Map Green image (\texttt{oid=1342213500}). Coordinates were taken from the new \textit{Herschel}/PACS Point Source Catalog.}
	\label{fig:hppsc2_example}
\end{figure*}

Figure~\ref{fig:hppsc2_example} shows examples of this functionality applied to real data. The top row depicts an original and processed image from the Scan Map Green observation with \texttt{oid=1342185553}. The left image in the first row shows the original input, and the right image displays the result after our tool has removed compact sources, leaving a background-only map. Similarly, the bottom row shows another example from Scan Map Green observation \texttt{oid=1342213500}. In both examples, the coordinates of compact sources were taken from the catalog by \citet{point_source_catalogue}. These examples demonstrate the tool’s effectiveness in removing cataloged compact sources from real \textit{Herschel} data, allowing users to analyze the underlying extended emission without interference from compact sources.

\subsection{Re-training on Custom Data}

While we include trained models for both Scan Map and Parallel Observing Modes, we also provide a small subset of our training data, enabling users to run the training pipeline and gain insight into its operation. The tool has been designed to allow for the inclusion of custom data sets, allowing users to implement their own data and train new models using our pipeline if required.

\section{Discussion}
\label{section:discussion}

In this section, we discuss our findings by presenting evaluation results of the model’s performance across different aspects. To ensure the reliability of our approach, we tested the model not only on simulated datasets but also on real sources by comparing the extracted fluxes with those of standard star models. This helped us confirm the accuracy of the extracted fluxes against well-calibrated reference values. We also analyzed the photometric accuracy on simulations to evaluate how well the model reconstructed the background and extracted sources. Additionally, we further assessed the preservation of image properties by evaluating the power spectrum and pixel intensity distribution of the resulting images. These evaluations provide a thorough overview of the model’s strengths and areas for improvement.

\subsection{Comparison with Standard Star Models}
\label{section:std_stars_comparison}

Studies by \citet{2014ExA....37..129B} and \citet{Klaas2018} have previously examined the flux calibration of the PACS photometer. We applied our network to estimate the background and extract the sources, then performed aperture photometry as discussed in Sect.~\ref{section:aperture_photometry} to measure the flux density of selected standard stars. We used the fluxes provided in these studies as references to evaluate both the accuracy of the \texttt{aperture\_photometry} function and the source removal performance of our network. Table~\ref{table:standard_stars} summarizes the comparison of the reference flux densities from the studies with the measured flux densities of the sources extracted by our network across the $70 \mu \mathrm{m}$, $100 \mu \mathrm{m}$, and $160 \mu \mathrm{m}$ bands.

These results validate that we can reliably use \texttt{aperture\_photometry} on the sources extracted by our network. Additionally, the extracted fluxes closely match the reference flux densities of the standard stars, as demonstrated by the flux ratios we recover in Table~\ref{table:standard_stars}. This confirms the ability of our model to accurately reconstruct real compact sources, ensuring minimal deviation from the established calibration values.

Some values are missing from the table due to limitations in the data retrieval process. We used the coordinates of the standard stars to download the maps from the \textit{Herschel} Science Archive website, performed a centroid fitting to refine the source positions, and then evaluated our model at those locations. However, in some cases, the calibration measurements were not processed up to at least Level2.5, preventing us from including them in our analysis. For the $160 \mu \mathrm{m}$ red band, some values at the bottom of the table are also missing because the sources were below the detection limit at that wavelength, making accurate flux measurements impossible.

\begin{table*}
    \caption{Comparison of PACS standard star reference flux densities with measured flux densities from extracted sources.}
    \label{table:standard_stars}
    \fontsize{7}{9}\selectfont
    \centering
    \begin{tabular}{ccc|cccc|cccc|cccc|c}
        \cline{4-15}
        \multicolumn{1}{l}{}                 & \multicolumn{1}{l}{}           & \multicolumn{1}{l|}{} & \multicolumn{4}{c|}{Scan Map Blue - 70 $\mu \mathrm{m}$}                                                           & \multicolumn{4}{c|}{Scan Map Green - 100 $\mu \mathrm{m}$}                                                         & \multicolumn{4}{c|}{Scan Map Red - 160 $\mu \mathrm{m}$}                                                           & \multicolumn{1}{l}{}     \\ \hline
        \multicolumn{1}{|c|}{name}           & \multicolumn{1}{c|}{ra}        & dec                   & \multicolumn{1}{c|}{ref.} & \multicolumn{1}{c|}{ref. err.} & \multicolumn{1}{c|}{ours}      & ratio & \multicolumn{1}{c|}{ref.} & \multicolumn{1}{c|}{ref. err.} & \multicolumn{1}{c|}{ours}      & ratio & \multicolumn{1}{c|}{ref.} & \multicolumn{1}{c|}{ref. err.} & \multicolumn{1}{c|}{ours}      & ratio & \multicolumn{1}{c|}{src.} \\ \hline
        \multicolumn{1}{|c|}{}               & \multicolumn{1}{c|}{{[}deg{]}} & {[}deg{]}             & \multicolumn{1}{c|}{{[}mJy{]}} & \multicolumn{1}{c|}{{[}mJy{]}}  & \multicolumn{1}{c|}{{[}mJy{]}} &       & \multicolumn{1}{c|}{{[}mJy{]}} & \multicolumn{1}{c|}{{[}mJy{]}}  & \multicolumn{1}{c|}{{[}mJy{]}} &       & \multicolumn{1}{c|}{{[}mJy{]}} & \multicolumn{1}{c|}{{[}mJy{]}}  & \multicolumn{1}{c|}{{[}mJy{]}} &       & \multicolumn{1}{c|}{}    \\ \hline
        \multicolumn{1}{|c|}{$\alpha$ Boo}   & \multicolumn{1}{c|}{213.915}   & 19.182                & \multicolumn{1}{c|}{15,434.0}  & \multicolumn{1}{c|}{}           & \multicolumn{1}{c|}{15,383.9} & 0.997 & \multicolumn{1}{c|}{7,509.0}   & \multicolumn{1}{c|}{}           & \multicolumn{1}{c|}{7,506.1}  & 1.000 & \multicolumn{1}{c|}{2,891.0}   & \multicolumn{1}{c|}{}           & \multicolumn{1}{c|}{2,900.5}  & 1.003 & \multicolumn{1}{c|}{(1)} \\
        \multicolumn{1}{|c|}{$\alpha$ Tau}   & \multicolumn{1}{c|}{68.980}    & 16.509                & \multicolumn{1}{c|}{14,131.0}  & \multicolumn{1}{c|}{}           & \multicolumn{1}{c|}{14,098.7} & 0.998 & \multicolumn{1}{c|}{6,909.0}   & \multicolumn{1}{c|}{}           & \multicolumn{1}{c|}{6,882.7}  & 0.996 & \multicolumn{1}{c|}{2,677.0}   & \multicolumn{1}{c|}{}           & \multicolumn{1}{c|}{2,677.5}  & 1.000 & \multicolumn{1}{c|}{(1)} \\
        \multicolumn{1}{|c|}{$\beta$ And}    & \multicolumn{1}{c|}{17.433}    & 35.621                & \multicolumn{1}{c|}{5,594.0}   & \multicolumn{1}{c|}{}           & \multicolumn{1}{c|}{5,608.5}  & 1.003 & \multicolumn{1}{c|}{2,737.0}   & \multicolumn{1}{c|}{}           & \multicolumn{1}{c|}{2,751.2}  & 1.005 & \multicolumn{1}{c|}{1,062.0}   & \multicolumn{1}{c|}{}           & \multicolumn{1}{c|}{1,068.0}  & 1.006 & \multicolumn{1}{c|}{(1)} \\
        \multicolumn{1}{|c|}{$\alpha$ Cet}   & \multicolumn{1}{c|}{45.570}    & 4.090                 & \multicolumn{1}{c|}{4,889.0}   & \multicolumn{1}{c|}{}           & \multicolumn{1}{c|}{4,884.9}  & 0.999 & \multicolumn{1}{c|}{2,393.0}   & \multicolumn{1}{c|}{}           & \multicolumn{1}{c|}{2,365.2}  & 0.988 & \multicolumn{1}{c|}{928.0}     & \multicolumn{1}{c|}{}           & \multicolumn{1}{c|}{925.0}    & 0.997 & \multicolumn{1}{c|}{(1)} \\
        \multicolumn{1}{|c|}{$\gamma$ Dra}   & \multicolumn{1}{c|}{269.152}   & 51.489                & \multicolumn{1}{c|}{3,283.0}   & \multicolumn{1}{c|}{}           & \multicolumn{1}{c|}{3,283.4}  & 1.000 & \multicolumn{1}{c|}{1,604.0}   & \multicolumn{1}{c|}{}           & \multicolumn{1}{c|}{1,607.8}  & 1.002 & \multicolumn{1}{c|}{621.0}     & \multicolumn{1}{c|}{}           & \multicolumn{1}{c|}{623.5}    & 1.004 & \multicolumn{1}{c|}{(1)} \\
        \multicolumn{1}{|c|}{$\beta$ Gem}    & \multicolumn{1}{c|}{116.329}   & 28.026                & \multicolumn{1}{c|}{2,457.0}   & \multicolumn{1}{c|}{140.8}    & \multicolumn{1}{c|}{2,634.1}  & 1.072 & \multicolumn{1}{c|}{1,190.0}   & \multicolumn{1}{c|}{68.2}     & \multicolumn{1}{c|}{1,226.7}  & 1.031 & \multicolumn{1}{c|}{455.9}     & \multicolumn{1}{c|}{26.1}     & \multicolumn{1}{c|}{468.7}    & 1.028 & \multicolumn{1}{c|}{(2)} \\
        \multicolumn{1}{|c|}{$\alpha$ Ari}   & \multicolumn{1}{c|}{31.793}    & 23.462                & \multicolumn{1}{c|}{1,707.0}   & \multicolumn{1}{c|}{100.7}    & \multicolumn{1}{c|}{1,705.6}  & 0.999 & \multicolumn{1}{c|}{831.4}     & \multicolumn{1}{c|}{49.1}     & \multicolumn{1}{c|}{833.5}    & 1.003 & \multicolumn{1}{c|}{321.0}     & \multicolumn{1}{c|}{18.9}     & \multicolumn{1}{c|}{323.1}    & 1.006 & \multicolumn{1}{c|}{(2)} \\
        \multicolumn{1}{|c|}{$\epsilon$ Lep} & \multicolumn{1}{c|}{76.365}    & -22.371               & \multicolumn{1}{c|}{1,182.0}   & \multicolumn{1}{c|}{69.7}     & \multicolumn{1}{c|}{1,185.0}  & 1.003 & \multicolumn{1}{c|}{576.2}     & \multicolumn{1}{c|}{34.0}     & \multicolumn{1}{c|}{578.6}    & 1.004 & \multicolumn{1}{c|}{222.7}     & \multicolumn{1}{c|}{13.1}     & \multicolumn{1}{c|}{213.8}    & 0.960 & \multicolumn{1}{c|}{(2)} \\
        \multicolumn{1}{|c|}{$\omega$ Cap}   & \multicolumn{1}{c|}{312.955}   & -26.919               & \multicolumn{1}{c|}{857.7}     & \multicolumn{1}{c|}{51.7}     & \multicolumn{1}{c|}{851.6}    & 0.993 & \multicolumn{1}{c|}{418.0}     & \multicolumn{1}{c|}{25.2}     & \multicolumn{1}{c|}{412.8}    & 0.988 & \multicolumn{1}{c|}{161.5}     & \multicolumn{1}{c|}{9.7}      & \multicolumn{1}{c|}{159.9}    & 0.990 & \multicolumn{1}{c|}{(2)} \\
        \multicolumn{1}{|c|}{$\eta$ Dra}     & \multicolumn{1}{c|}{245.998}   & 61.514                & \multicolumn{1}{c|}{479.5}     & \multicolumn{1}{c|}{16.2}    & \multicolumn{1}{c|}{515.8}    & 1.076 & \multicolumn{1}{c|}{232.6}     & \multicolumn{1}{c|}{8.0}      & \multicolumn{1}{c|}{220.7}    & 0.949 & \multicolumn{1}{c|}{89.4}      & \multicolumn{1}{c|}{3.1}      & \multicolumn{1}{c|}{83.0}     & 0.928 & \multicolumn{1}{c|}{(2)} \\
        \multicolumn{1}{|c|}{$\delta$ Dra}   & \multicolumn{1}{c|}{288.139}   & 67.662                & \multicolumn{1}{c|}{428.9}     & \multicolumn{1}{c|}{24.5}     & \multicolumn{1}{c|}{436.9}    & 1.019 & \multicolumn{1}{c|}{207.7}     & \multicolumn{1}{c|}{11.8}     & \multicolumn{1}{c|}{210.9}    & 1.015 & \multicolumn{1}{c|}{79.6}      & \multicolumn{1}{c|}{4.5}      & \multicolumn{1}{c|}{85.0}     & 1.067 & \multicolumn{1}{c|}{(2)} \\
        \multicolumn{1}{|c|}{$\theta$ Umi}   & \multicolumn{1}{c|}{232.854}   & 77.349                & \multicolumn{1}{c|}{286.2}     & \multicolumn{1}{c|}{16.2}     & \multicolumn{1}{c|}{288.5}    & 1.008 & \multicolumn{1}{c|}{139.5}     & \multicolumn{1}{c|}{7.9}      & \multicolumn{1}{c|}{}         &      & \multicolumn{1}{c|}{53.9}      & \multicolumn{1}{c|}{3.1}      & \multicolumn{1}{c|}{}         &      & \multicolumn{1}{c|}{(2)} \\
        \multicolumn{1}{|c|}{HD 41047}       & \multicolumn{1}{c|}{90.318}    & -33.912               & \multicolumn{1}{c|}{195.6}     & \multicolumn{1}{c|}{11.7}     & \multicolumn{1}{c|}{}         &      & \multicolumn{1}{c|}{95.4}      & \multicolumn{1}{c|}{5.7}      & \multicolumn{1}{c|}{95.7}     & 1.003 & \multicolumn{1}{c|}{36.9}      & \multicolumn{1}{c|}{2.2}      & \multicolumn{1}{c|}{28.2}     & 0.764 & \multicolumn{1}{c|}{(2)} \\
        \multicolumn{1}{|c|}{42 Dra}         & \multicolumn{1}{c|}{276.496}   & 65.563                & \multicolumn{1}{c|}{153.7}     & \multicolumn{1}{c|}{4.6}        & \multicolumn{1}{c|}{149.1}    & 0.970 & \multicolumn{1}{c|}{75.3}      & \multicolumn{1}{c|}{2.3}      & \multicolumn{1}{c|}{72.1}     & 0.957 & \multicolumn{1}{c|}{29.4}      & \multicolumn{1}{c|}{0.9}      & \multicolumn{1}{c|}{30.1}     & 1.023 & \multicolumn{1}{c|}{(2)} \\
        \multicolumn{1}{|c|}{HD 138265}      & \multicolumn{1}{c|}{231.964}   & 60.670                & \multicolumn{1}{c|}{115.9}     & \multicolumn{1}{c|}{4.0}        & \multicolumn{1}{c|}{113.8}    & 0.982 & \multicolumn{1}{c|}{56.8}      & \multicolumn{1}{c|}{2.0}      & \multicolumn{1}{c|}{55.8}     & 0.982 & \multicolumn{1}{c|}{22.2}      & \multicolumn{1}{c|}{0.8}      & \multicolumn{1}{c|}{20.5}     & 0.925 & \multicolumn{1}{c|}{(2)} \\
        \multicolumn{1}{|c|}{HD 159330}      & \multicolumn{1}{c|}{262.682}   & 57.877                & \multicolumn{1}{c|}{64.2}      & \multicolumn{1}{c|}{2.1}        & \multicolumn{1}{c|}{62.2}     & 0.969 & \multicolumn{1}{c|}{31.5}      & \multicolumn{1}{c|}{1.0}      & \multicolumn{1}{c|}{31.3}     & 0.995 & \multicolumn{1}{c|}{12.3}      & \multicolumn{1}{c|}{0.4}      & \multicolumn{1}{c|}{12.7}     & 1.033 & \multicolumn{1}{c|}{(2)} \\
        \multicolumn{1}{|c|}{HD 152222}      & \multicolumn{1}{c|}{251.769}   & 67.267                & \multicolumn{1}{c|}{39.4}      & \multicolumn{1}{c|}{1.9}        & \multicolumn{1}{c|}{40.7}     & 1.034 & \multicolumn{1}{c|}{19.3}      & \multicolumn{1}{c|}{1.0}      & \multicolumn{1}{c|}{20.6}     & 1.065 & \multicolumn{1}{c|}{7.5}       & \multicolumn{1}{c|}{0.1}      & \multicolumn{1}{c|}{}         &      & \multicolumn{1}{c|}{(2)} \\
        \multicolumn{1}{|c|}{HD 39608}       & \multicolumn{1}{c|}{87.402}    & -60.676               & \multicolumn{1}{c|}{30.9}      & \multicolumn{1}{c|}{1.2}        & \multicolumn{1}{c|}{30.6}     & 0.990 & \multicolumn{1}{c|}{15.1}      & \multicolumn{1}{c|}{0.6}      & \multicolumn{1}{c|}{}         &      & \multicolumn{1}{c|}{5.9}       & \multicolumn{1}{c|}{0.2}      & \multicolumn{1}{c|}{}         &      & \multicolumn{1}{c|}{(2)} \\
        \multicolumn{1}{|c|}{$\delta$ Hyi}   & \multicolumn{1}{c|}{35.437}    & -68.659               & \multicolumn{1}{c|}{22.9}      & \multicolumn{1}{c|}{0.8}        & \multicolumn{1}{c|}{20.9}     & 0.914 & \multicolumn{1}{c|}{11.2}      & \multicolumn{1}{c|}{0.4}      & \multicolumn{1}{c|}{10.3}     & 0.920 & \multicolumn{1}{c|}{4.4}       & \multicolumn{1}{c|}{0.2}      & \multicolumn{1}{c|}{}         &      & \multicolumn{1}{c|}{(2)} \\ \hline
    \end{tabular}
    \tablefoot{Columns include star name, J2000 coordinates, reference and measured flux densities at $70 \mu \mathrm{m}$, $100 \mu \mathrm{m}$, and $160 \mu \mathrm{m}$ with errors (when available), their ratios, and the source of reference values. References are (1) \cite{2014ExA....37..129B} and (2) \cite{Klaas2018}.}
\end{table*}

\subsection{Photometric Accuracy}

As described in Sect.~\ref{section:train_test_split}, we preserved several fields for testing purposes, with the numbers shown in Table~\ref{table:sources_and_fields}. These fields were not used in either the training or validation loops, making them suitable for testing our model on unseen cases.

An important aspect of evaluating our model’s performance involves measuring the flux densities of the extracted sources and comparing them to the theoretical flux densities of the injected PSFs. If the measured flux density of the extracted source closely aligns with the known value, it indicates that our model has accurately estimated the background and successfully removed the source with precision.

We applied the same processing steps to our test dataset as we did to the other datasets: each image was per-image normalized, mask sizes were determined using the same parameters, and all inputs were passed through our pre-trained model to create source-free, background-only images for all injected sources. We obtained the extracted source by subtracting the model's predicted background from the original input image. Next, we performed aperture photometry on this background-subtracted source with various aperture sizes, ranging from 4 to 18 arcseconds, to measure the extracted flux density. We also performed circular annulus background estimation on the input images, as described in Sect.~\ref{section:aperture_photometry}, and compared these flux density values to the known injected flux densities by calculating their ratios.

\begin{figure*}
	\resizebox{\hsize}{!}
	{\includegraphics[width=\hsize]{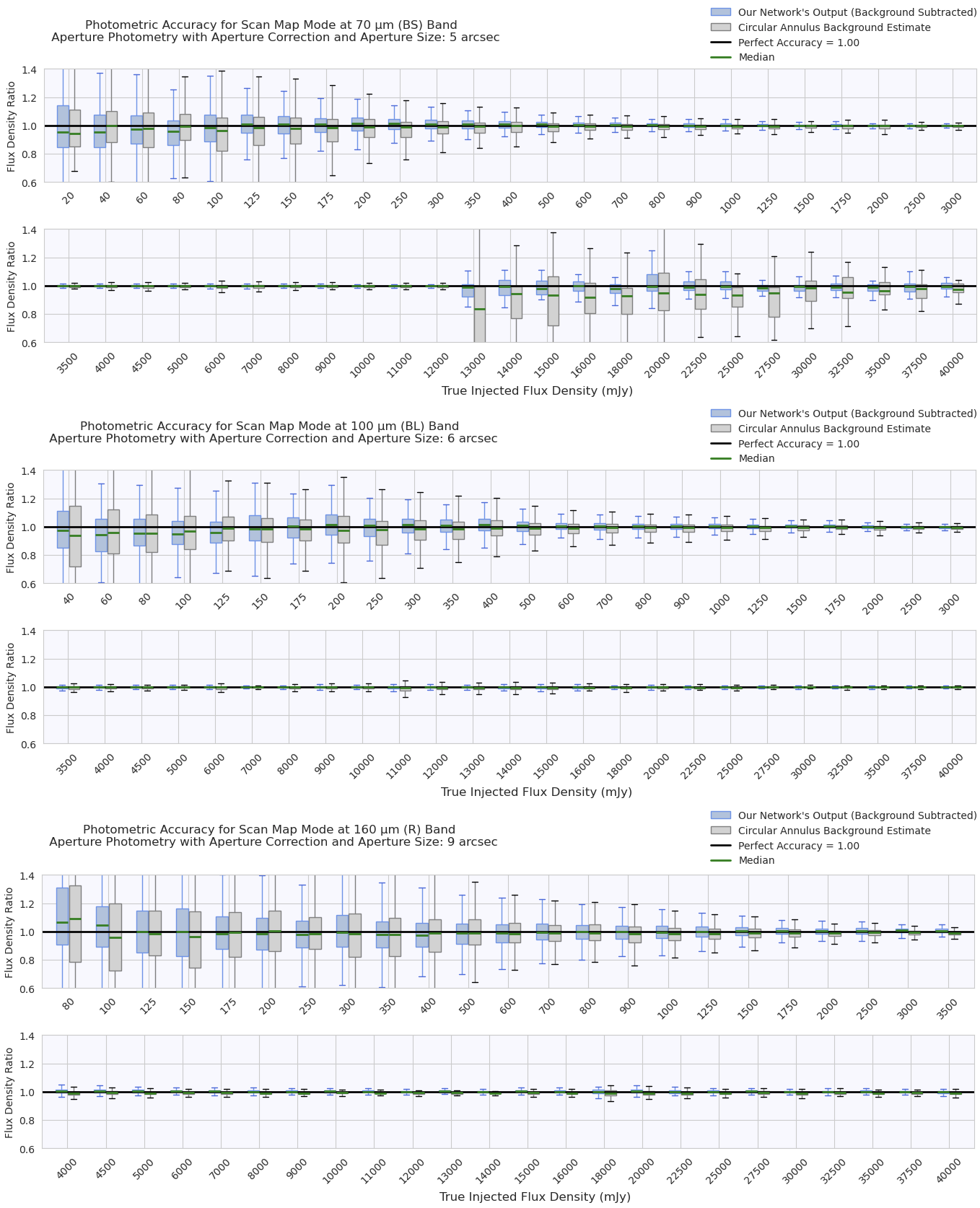}}
	\caption{Photometric accuracy evaluation for the Scan Map mode across three bands. Top: Scan Map blue ($70 \mu \mathrm{m}$ - BS band); Middle: Scan Map green ($100 \mu \mathrm{m}$ - BL band); Bottom: Scan Map red ($160 \mu \mathrm{m}$ - R band). Blue boxes represent the flux density ratios of sources extracted by our model to the known injected fluxes, while gray boxes show ratios using standard circular annulus background estimation without source removal. The line at $y=1$ indicates perfect flux matching.}
	\label{fig:photometry_scanmap}
\end{figure*}

\begin{figure*}
	\resizebox{\hsize}{!}
	{\includegraphics[width=\hsize]{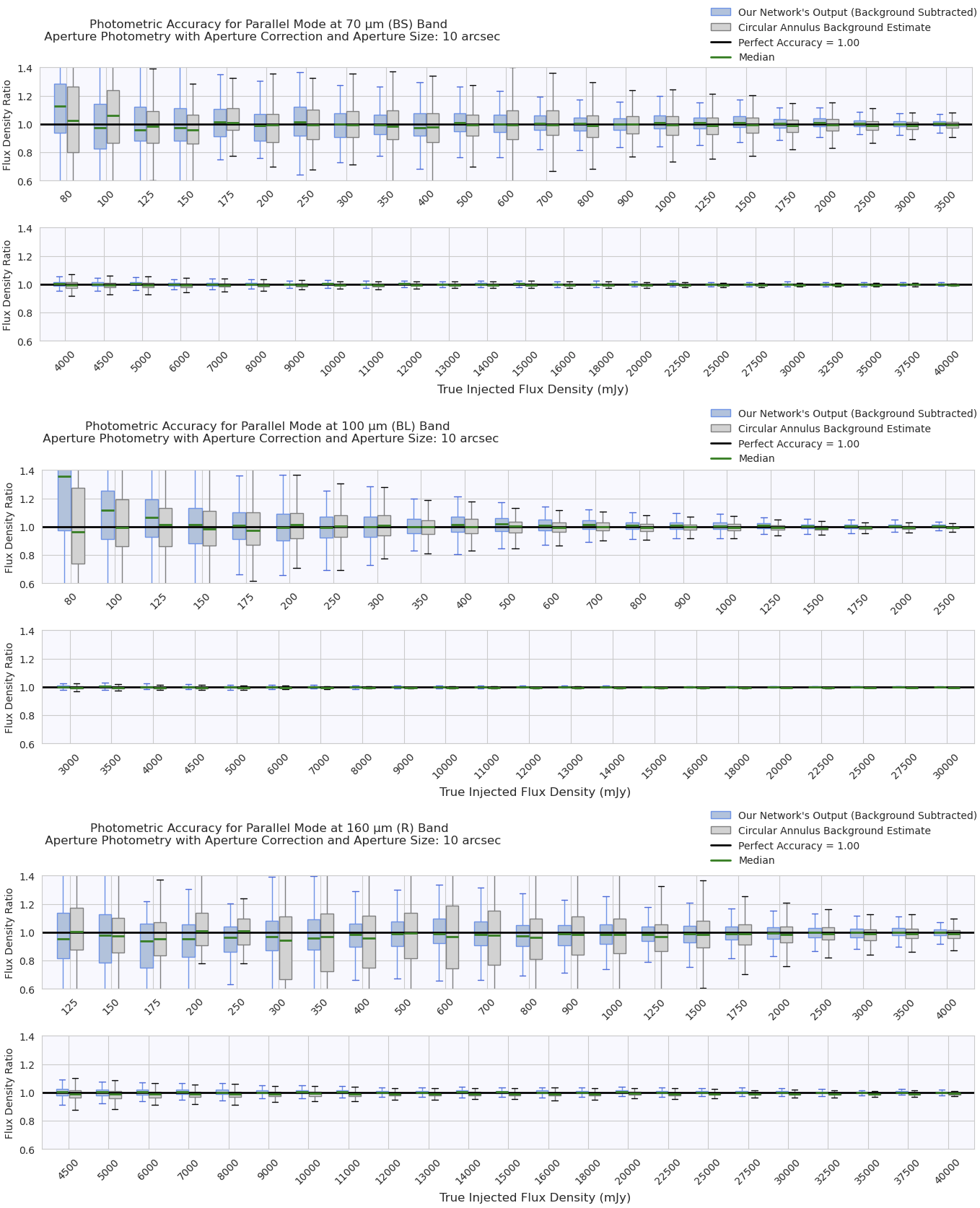}}
	\caption{Photometric accuracy evaluation for the Parallel mode across three bands. Top: Parallel blue (\(70 \mu \mathrm{m}\) - BS band); Middle: Parallel green (\(100 \mu \mathrm{m}\) - BL band); Bottom: Parallel red (\(160 \mu \mathrm{m}\) - R band).}
	\label{fig:photometry_parallel}
\end{figure*}

Figure~\ref{fig:photometry_scanmap} shows the results for the Scan Map mode, while Fig.~\ref{fig:photometry_parallel} illustrates results for the Parallel mode. In these figures, blue box plots represent the ratios of the flux densities measured on sources extracted by our model to the known injected flux densities. Additionally, gray box plots show the flux ratios measured directly from the input images using circular annulus background estimation, without model-based source removal. The horizontal line at $y=1$ indicates the ideal flux ratio, where the measured flux exactly matches the known value. While the gray boxes reflect the flux density estimations from standard photometry, the blue boxes assess the model’s effectiveness in accurately estimating the background by removing the injected source. A flux ratio of 1 indicates that the model has precisely removed the injected flux, achieving an accurate background estimate.

Two key results are visible in the figures. First, our model accurately removes the sources with high precision, as seen in the close alignment of the green median lines in the blue boxes with the ideal line. Second, not only is the source removal accurate on average, but for most cases, the flux estimates for the extracted sources are more consistent than those obtained with circular annulus background estimation, as shown in the narrower blue box plots. This reduced interquartile range (IQR) suggests that the model offers a more stable and reliable flux measurements with less variation around the median compared to the annulus method.

It is worth noting that the IQR is generally larger for sources with lower flux densities. This is because we set the lowest flux levels for all fields where sources only begin to appear against the background, leading to larger deviations from the ideal as these flux levels approach the detection threshold. Also, in the Scan Map $70 \mu \mathrm{m}$ (BS) band plot, starting at around $13,000$ mJy, a noticeable increase in deviation can be observed, which is due to sources in regions with particularly complex backgrounds, such as dense clouds and complex structures. Despite these challenges, the model-based approach generally provides more accurate background estimates than the circular annulus method in these difficult environments.

\subsection{Image comparison}

To evaluate the performance of our method, we further compare the properties of the original source-free images with the source-free predictions obtained after removing simulated sources. In this subsection, we present two analyses to assess the alignment between these maps: the power spectrum and the pixel intensity distribution. The aim of these approaches is to ensure that the predicted maps preserve the structural and statistical characteristics of the original source-free images, demonstrating the effectiveness of the source removal process.

The power spectrum analysis examines the spatial frequency domain, providing insights into whether the model has preserved the underlying structures and avoided introducing artifacts. Meanwhile, the pixel intensity distribution analysis evaluates the statistical properties of pixel values in the predicted maps, ensuring that the overall image characteristics remain consistent with the original. Together, these evaluations provide a great assessment of the method’s ability to accurately reproduce the original source-free image properties.

\begin{figure*}[ht]
	\resizebox{\hsize}{!}
	{\includegraphics[width=\hsize]{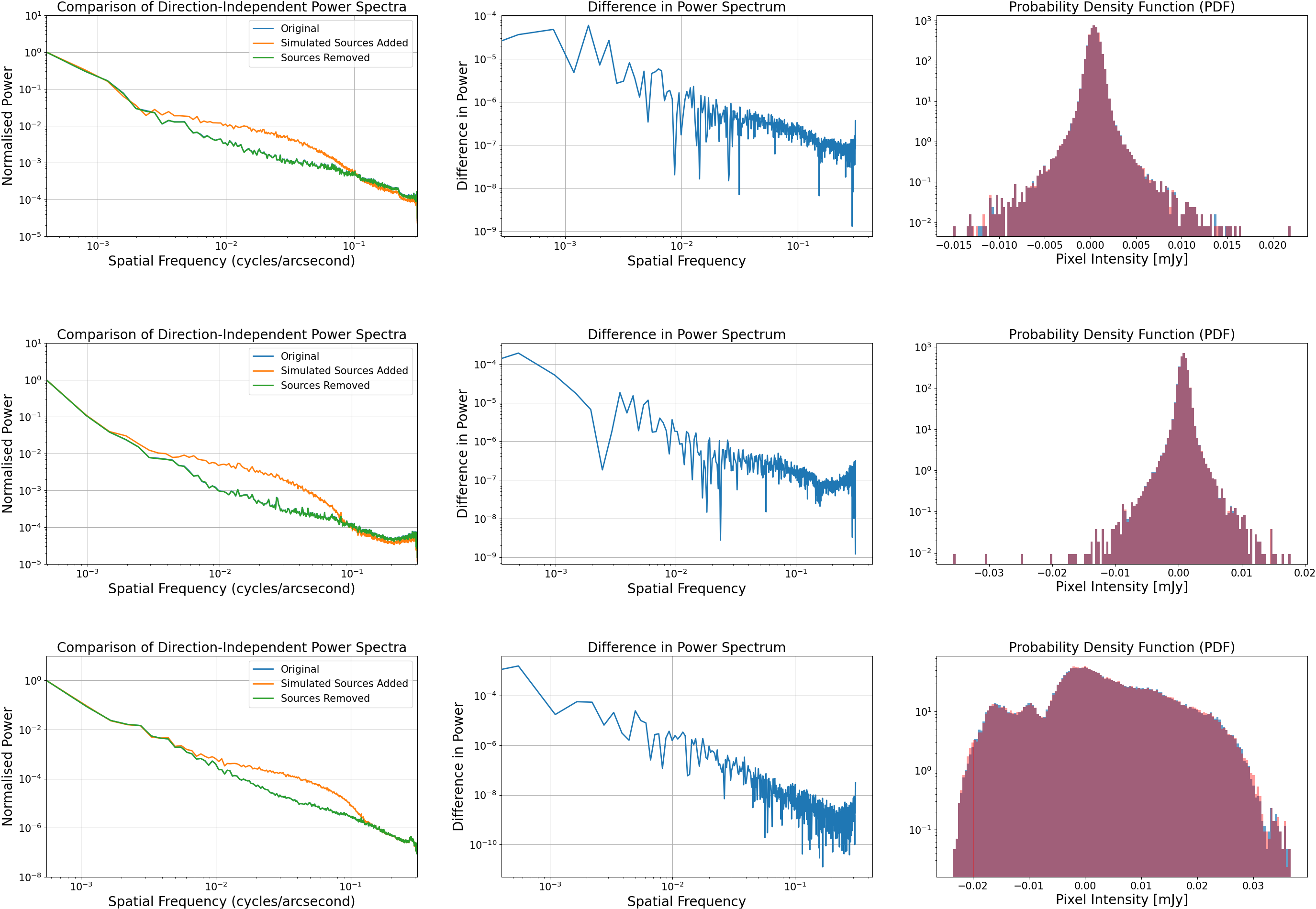}}
\caption{\revision{From top to bottom, the three rows correspond to the Scan Map Blue, Green, and Red fields, with observation IDs \texttt{1342216409}, \texttt{1342244224}, \texttt{1342249136}, respectively.} Left column: Direction independent Fourier transform power spectra of \revision{each field} before adding artificial sources (blue line), with the artificial sources (orange) and after their removal (green). Power spectra are normalized. 
Middle column: Difference of the power spectrum before adding sources and after their removal \revision{for each field}. 
Right column: \revision{Probability density functions before adding artificial sources (blue histograms) and after their removal (red histograms).}}\label{fig:ps}   
\end{figure*}

\subsubsection{Power spectrum}
The Fourier transform power spectra of the different maps provide useful insight into the properties of the signal introduced by our tool. In Fig.~\ref{fig:ps} we demonstrate a comparison with one case in each filter band \revision{in the Scan Map dataset}. These fields were chosen from the test dataset, and for each field, we generated random positions for the sources. For each random position, a random flux density and orientation were also assigned, ensuring that the fields represent the most diverse and variable cases possible. These maps are similar to the example shown in Fig.~\ref{fig:sim_result_example}.

The images used for this comparison are as follows: Scan Map Blue \texttt{oid=1342216409}, Scan Map Green \texttt{oid=1342244224}, and Scan Map Red \texttt{oid=1342249136}. This order corresponds to the arrangement of the bands from top to bottom in Fig.~\ref{fig:ps}.

Plots in the left hand column show the power spectra of the fields. They clearly illustrate that adding the artificial sources significantly modifies the power spectrum (orange lines), especially at the spatial scale similar to the size of the point sources. They also clearly show that the original (blue lines) and the predicted maps (green lines) are very close and actually cannot be distinguished on these plots, the plotting lines do completely overlap. To highlight any subtle differences, we also plotted the difference between the power spectra of the original and predicted maps, shown in the middle column. These plots reveal that the differences in the normalized spectra are only on the order of 10$^{-4}$ to 10$^{-10}$, depending on the spatial frequency.

%

\subsubsection{Pixel intensity distribution}

As an additional test, we examined the PDF of the fields. Examples are shown in the right column of Fig.~\ref{fig:ps}, focusing on the masked regions where compact sources were injected and subsequently removed. The PDFs of the original source-free maps (blue histograms) and the predicted source-free maps (red histograms) show strong agreement. This demonstrates that the model accurately reconstructs the extended emission and preserves the statistical properties of the pixel intensity distribution, including noise features and background variations.

Table~\ref{table:metrics_complete} provides a detailed comparison of pixel intensity statistics in the masked regions. Metrics such as the mean, median, standard deviation, skewness, and interquartile range were analyzed. The alignment of mean and median values between the original and predicted maps indicates that the model effectively recovers the background emission without introducing biases in brightness. Furthermore, the agreement in standard deviation and IQR confirms that the model preserves the spread and variability of the background intensity. The consistency in skewness further highlights the model’s ability to maintain asymmetry in the pixel intensity distribution, reflecting complex structures in the interstellar medium.

These results validate that the model not only removes compact sources across a range of flux densities but also preserves the underlying extended emission and statistical integrity of the images. This consistency across all test fields, observing modes, and filter bands delivers reliable maps for subsequent astrophysical analysis.

\afterpage{
\begin{landscape} 
\vspace*{\fill} 
\begin{table}[htbp]
    \centering
    \caption{Comparison of pixel intensity statistics between original and predicted maps across test fields.}
    \label{table:metrics_complete}
    \renewcommand{\arraystretch}{1.5} 
    \footnotesize
    \begin{tabular}{|c|c|c|c|c|c|c|c|c|c|}
        \hline
Type                           & OBSID & Original Mean & Predicted Mean  & Original Median & Predicted Median & Original  & Predicted & Original & Predicted \\ 
& &  $\pm$ Stdev & $\pm$ Stdev & $\pm$ MAD &  $\pm$ MAD& Skewness &Skewness & IQR & IQR \\
\hline
\multirow{3}{*}{Scan Map Blue}  & 1342216409 & $4.07\mathrm{e}{-4} \pm 6.47\mathrm{e}{-4}$ & $4.12\mathrm{e}{-4} \pm 6.45\mathrm{e}{-4}$ & $4.08\mathrm{e}{-4} \pm 3.58\mathrm{e}{-4}$ & $4.12\mathrm{e}{-4} \pm 3.54\mathrm{e}{-4}$ & $1.17\mathrm{e}{-1}$ & $1.19\mathrm{e}{-1}$ & $7.15\mathrm{e}{-4}$ & $7.09\mathrm{e}{-4}$ \\ \cline{2-10} 
                                & 1342237154 & $1.59\mathrm{e}{-3} \pm 9.80\mathrm{e}{-3}$ & $1.61\mathrm{e}{-3} \pm 9.81\mathrm{e}{-3}$ & $-3.31\mathrm{e}{-5} \pm 6.01\mathrm{e}{-3}$ & $-1.82\mathrm{e}{-5} \pm 6.01\mathrm{e}{-3}$ & $1.03\mathrm{e}{0}$ & $1.03\mathrm{e}{0}$ & $1.22\mathrm{e}{-2}$ & $1.21\mathrm{e}{-2}$ \\ \cline{2-10} 
                               & 1342249954 & $2.26\mathrm{e}{-3} \pm 1.47\mathrm{e}{-2}$ & $2.30\mathrm{e}{-3} \pm 1.47\mathrm{e}{-2}$ & $-1.16\mathrm{e}{-4} \pm 8.47\mathrm{e}{-3}$ & $-8.08\mathrm{e}{-5} \pm 8.47\mathrm{e}{-3}$ & $1.06\mathrm{e}{0}$ & $1.06\mathrm{e}{0}$ & $1.75\mathrm{e}{-2}$ & $1.75\mathrm{e}{-2}$ \\ \hline
\multirow{3}{*}{Scan Map Green}& 1342189388 & $1.67\mathrm{e}{-2} \pm 5.05\mathrm{e}{-2}$ & $1.68\mathrm{e}{-2} \pm 5.05\mathrm{e}{-2}$ & $8.43\mathrm{e}{-4} \pm 4.57\mathrm{e}{-3}$ & $8.99\mathrm{e}{-4} \pm 4.55\mathrm{e}{-3}$ & $5.92\mathrm{e}{0}$ & $5.93\mathrm{e}{0}$ & $1.13\mathrm{e}{-2}$ & $1.13\mathrm{e}{-2}$ \\ \cline{2-10} 
                                & 1342217764 & $1.24\mathrm{e}{-2} \pm 4.71\mathrm{e}{-2}$ & $1.25\mathrm{e}{-2} \pm 4.71\mathrm{e}{-2}$ & $0.0 \pm 1.80\mathrm{e}{-2}$ & $-1.59\mathrm{e}{-5} \pm 1.78\mathrm{e}{-2}$ & $2.32\mathrm{e}{0}$ & $2.32\mathrm{e}{0}$ & $3.70\mathrm{e}{-2}$ & $3.68\mathrm{e}{-2}$ \\ \cline{2-10} 
                               & 1342244224 & $7.57\mathrm{e}{-4} \pm 7.87\mathrm{e}{-4}$ & $7.67\mathrm{e}{-4} \pm 7.85\mathrm{e}{-4}$ & $7.69\mathrm{e}{-4} \pm 3.83\mathrm{e}{-4}$ & $7.79\mathrm{e}{-4} \pm 3.80\mathrm{e}{-4}$ & $-1.38\mathrm{e}{0}$ & $-1.40\mathrm{e}{0}$ & $7.66\mathrm{e}{-4}$ & $7.60\mathrm{e}{-4}$ \\ \hline
\multirow{3}{*}{Scan Map Red}  & 1342204168 & $9.67\mathrm{e}{-5} \pm 5.87\mathrm{e}{-3}$ & $7.68\mathrm{e}{-5} \pm 5.89\mathrm{e}{-3}$ & $-7.95\mathrm{e}{-4} \pm 4.28\mathrm{e}{-3}$ & $-8.19\mathrm{e}{-4} \pm 4.32\mathrm{e}{-3}$ & $4.34\mathrm{e}{-1}$ & $4.23\mathrm{e}{-1}$ & $8.70\mathrm{e}{-3}$ & $8.77\mathrm{e}{-3}$ \\ \cline{2-10} 
                                & 1342249136 & $2.91\mathrm{e}{-3} \pm 9.88\mathrm{e}{-3}$ & $2.92\mathrm{e}{-3} \pm 9.87\mathrm{e}{-3}$ & $1.83\mathrm{e}{-3} \pm 5.72\mathrm{e}{-3}$ & $1.82\mathrm{e}{-3} \pm 5.67\mathrm{e}{-3}$ & $1.43\mathrm{e}{-1}$ & $1.44\mathrm{e}{-1}$ & $1.22\mathrm{e}{-2}$ & $1.21\mathrm{e}{-2}$ \\ \cline{2-10} 
                               & 1342249269 & $1.85\mathrm{e}{-2} \pm 8.71\mathrm{e}{-2}$ & $1.87\mathrm{e}{-2} \pm 8.70\mathrm{e}{-2}$ & $1.11\mathrm{e}{-2} \pm 5.63\mathrm{e}{-2}$ & $1.14\mathrm{e}{-2} \pm 5.62\mathrm{e}{-2}$ & $8.73\mathrm{e}{-1}$ & $8.70\mathrm{e}{-1}$ & $1.13\mathrm{e}{-1}$ & $1.12\mathrm{e}{-1}$ \\ \hline

\multirow{3}{*}{Parallel Blue}  & 1342185644 & $-6.76\mathrm{e}{-4} \pm 2.40\mathrm{e}{-3}$ & $-6.73\mathrm{e}{-4} \pm 2.39\mathrm{e}{-3}$ & $-6.88\mathrm{e}{-4} \pm 1.60\mathrm{e}{-3}$ & $-6.67\mathrm{e}{-4} \pm 1.49\mathrm{e}{-3}$ & $2.64\mathrm{e}{-2}$ & $3.54\mathrm{e}{-2}$ & $3.19\mathrm{e}{-3}$ & $2.99\mathrm{e}{-3}$ \\ \cline{2-10} 
                                & 1342244170 & $6.42\mathrm{e}{-3} \pm 3.17\mathrm{e}{-2}$ & $6.50\mathrm{e}{-4} \pm 2.31\mathrm{e}{-3}$ & $-1.78\mathrm{e}{-3} \pm 1.67\mathrm{e}{-2}$ & $-1.76\mathrm{e}{-3} \pm 1.66\mathrm{e}{-2}$ & $1.14\mathrm{e}{0}$ & $1.13\mathrm{e}{0}$ & $3.83\mathrm{e}{-2}$ & $3.78\mathrm{e}{-2}$ \\ \cline{2-10} 
                                & 1342257384 & $5.44\mathrm{e}{-3} \pm 1.15\mathrm{e}{-2}$ & $5.61\mathrm{e}{-3} \pm 1.14\mathrm{e}{-2}$ & $3.47\mathrm{e}{-3} \pm 6.98\mathrm{e}{-3}$ & $3.62\mathrm{e}{-3} \pm 6.99\mathrm{e}{-3}$ & $9.12\mathrm{e}{-1}$ & $9.08\mathrm{e}{-1}$ & $1.45\mathrm{e}{-2}$ & $1.45\mathrm{e}{-2}$ \\ \hline
\multirow{2}{*}{Parallel Green} & 1342197280 & $1.85\mathrm{e}{-2} \pm 1.97\mathrm{e}{-3}$ & $1.86\mathrm{e}{-2} \pm 1.92\mathrm{e}{-3}$ & $1.85\mathrm{e}{-2} \pm 1.32\mathrm{e}{-3}$ & $1.86\mathrm{e}{-2} \pm 1.26\mathrm{e}{-3}$ & $3.06\mathrm{e}{-3}$ & $-1.18\mathrm{e}{-2}$ & $2.65\mathrm{e}{-3}$ & $2.53\mathrm{e}{-3}$ \\ \cline{2-10} 
                                & 1342239263 & $5.39\mathrm{e}{-3} \pm 2.82\mathrm{e}{-3}$ & $5.43\mathrm{e}{-3} \pm 2.80\mathrm{e}{-3}$ & $5.23\mathrm{e}{-3} \pm 1.89\mathrm{e}{-3}$ & $5.29\mathrm{e}{-3} \pm 1.86\mathrm{e}{-3}$ & $2.58\mathrm{e}{-1}$ & $2.59\mathrm{e}{-1}$ & $3.79\mathrm{e}{-3}$ & $3.73\mathrm{e}{-3}$ \\ \hline
\multirow{3}{*}{Parallel Red}  & 1342185644 & $-4.42\mathrm{e}{-4} \pm 1.91\mathrm{e}{-3}$ & $-4.51\mathrm{e}{-4} \pm 1.87\mathrm{e}{-3}$ & $-4.53\mathrm{e}{-4} \pm 1.26\mathrm{e}{-3}$ & $-4.73\mathrm{e}{-4} \pm 1.23\mathrm{e}{-3}$ & $4.33\mathrm{e}{-2}$ & $6.85\mathrm{e}{-2}$ & $2.52\mathrm{e}{-3}$ & $2.47\mathrm{e}{-3}$ \\ \cline{2-10} 
                               & 1342186121 & $4.10\mathrm{e}{-2} \pm 4.80\mathrm{e}{-2}$ & $4.11\mathrm{e}{-2} \pm 4.79\mathrm{e}{-2}$ & $3.42\mathrm{e}{-2} \pm 1.99\mathrm{e}{-2}$ & $3.43\mathrm{e}{-2} \pm 1.99\mathrm{e}{-2}$ & $1.03\mathrm{e}{1}$ & $1.03\mathrm{e}{1}$ & $4.04\mathrm{e}{-2}$ & $4.05\mathrm{e}{-2}$ \\ \cline{2-10} 
                               & 1342257382 & $9.32\mathrm{e}{-2} \pm 1.03\mathrm{e}{-1}$ & $9.33\mathrm{e}{-2} \pm 1.03\mathrm{e}{-1}$ & $6.79\mathrm{e}{-2} \pm 4.75\mathrm{e}{-2}$ & $6.81\mathrm{e}{-2} \pm 4.74\mathrm{e}{-2}$ & $8.11\mathrm{e}{0}$ & $8.16\mathrm{e}{0}$ & $1.07\mathrm{e}{-1}$ & $1.07\mathrm{e}{-1}$ \\ \hline
    \end{tabular}
    \tablefoot{Units are mJy/pixel}
\end{table}
\vspace*{\fill} 
\end{landscape} 
}

\section{Summary}

We introduced a deep learning approach for effective source removal from photometric observations. The goal of the method was two-fold: analysis of the extended emission is less effective in the presence of point sources, as their presence is disturbing to the power spectrum and other image analysis techniques. On the other hand, the fluctuating background is a problem for many photometry methods, as the background cannot be easily estimated. 

To address these challenges, we presented a novel source removal method tailored for photometric observations made by the \textit{Herschel}/PACS instrument. This approach was based on a U-Net-like architecture enhanced by Partial Convolution blocks. It integrated carefully selected loss functions, including L1 loss, style loss, SSIM loss, and custom source loss, ensuring seamless background reconstruction and accurate source extraction. The use of encoder-decoder symmetry with skip connections effectively preserved spatial information while extracting complex features from the input data. A dynamically calculated mask further improved the efficiency and accuracy of the process. Key design elements included the Partial Convolution Downsample and Upsample blocks, tailored for handling masked inputs and achieving efficient inpainting of source-free regions.

The performance of the architecture was evaluated using multiple approaches. First, we assessed background reconstruction in the masked areas by comparing the amount of injected and removed flux, demonstrating that our method provided more consistent results for aperture photometry compared to the standard approach, which lacked robust background estimation. We also compared the properties of the original and predicted images through Fourier transform power spectra and probability density functions, finding that the original image characteristics were preserved across all spatial scales and intensity distributions. Additionally, we validated the accuracy of our method on real sources by comparing extracted fluxes to reference values from standard star models, confirming its reliability on real data.

In conclusion, the proposed approach offers a powerful tool for the community, significantly reducing manual efforts in image processing while enhancing the reliability of background estimations and source extractions. Although the method was applied to \textit{Herschel}/PACS observations, it can be trained on other photometric maps as well. We also provided a Python package with tutorials and examples for the users.

\begin{acknowledgements}
  This work has received funding from the European Union's Horizon 2020 research and innovation program under grant number 101004141. G.M. acknowledges support from the János Bolyai Research Scholarship of the Hungarian Academy of Sciences. This research was supported by the International Space Science Institute (ISSI) in Bern, through ISSI International Team project 521 selected in 2021, Revisiting Star Formation in the Era of Big Data (\url{https://teams.issibern.ch/starformation/}). \revision{We are grateful to our referee for the insightful comments and suggestions that have greatly contributed to further improving this paper.}
\end{acknowledgements}

\bibliographystyle{aa}
\bibliography{references}

\begin{appendix}
    \section{Details of the Simulations}

\begin{table*}[]
    \caption{Details of the dataset for the Scan Map Observing Mode in the Blue band.}
    \label{table:scanmap_blue_details}
    \centering
    \begin{tabular}{|c|c|c|c|c|c|c|}
    \hline
        \textbf{Field}    & \textbf{\# Sources} & \textbf{Min Flux} & \textbf{Max Flux} & \textbf{Dataset} & \textbf{Total Sources} & \textbf{\# Simulations} \\
        \hline
        1342227809\_L2\_5 & 190                 & 700               & 30000             & Training         & 5890                   & 31                      \\
        1342195485\_L2\_5 & 170                 & 20                & 6000              & Training         & 5100                   & 30                      \\
        1342216487\_L2\_5 & 170                 & 150               & 15000             & Training         & 5610                   & 33                      \\
        1342258816\_L2\_5 & 150                 & 150               & 14000             & Training         & 4800                   & 32                      \\
        1342191945\_L2\_5 & 145                 & 10                & 5000              & Training         & 4350                   & 30                      \\
        1342241633\_L2\_5 & 120                 & 10                & 6000              & Training         & 3720                   & 31                      \\
        1342188474\_L2\_5 & 100                 & 10                & 3500              & Training         & 2700                   & 27                      \\
        1342246718\_L2\_5 & 100                 & 20                & 8000              & Training         & 3200                   & 32                      \\
        1342191106\_L2\_5 & 90                  & 2000              & 40000             & Training         & 2520                   & 28                      \\
        1342183905\_L3    & 70                  & 10                & 4000              & Training         & 1960                   & 28                      \\
        1342187003\_L2\_5 & 70                  & 40                & 6000              & Training         & 2030                   & 29                      \\
        1342191805\_L2\_5 & 70                  & 80                & 14000             & Training         & 2450                   & 35                      \\
        1342191801\_L2\_5 & 70                  & 80                & 15000             & Training         & 2520                   & 36                      \\
        1342187077\_L2\_5 & 60                  & 60                & 6000              & Training         & 1680                   & 28                      \\
        1342237428\_L2\_5 & 60                  & 20                & 7000              & Training         & 1860                   & 31                      \\
        1342217754\_L2\_5 & 55                  & 40                & 8000              & Training         & 1705                   & 31                      \\
        1342217752\_L2\_5 & 55                  & 200               & 35000             & Training         & 2200                   & 40                      \\
        1342217758\_L2\_5 & 55                  & 250               & 30000             & Training         & 2035                   & 37                      \\
        1342228952\_L2\_5 & 55                  & 20                & 8000              & Training         & 1760                   & 32                      \\
        1342190339\_L2\_5 & 50                  & 10                & 4000              & Training         & 1400                   & 28                      \\
        1342209668\_L2\_5 & 50                  & 20                & 4000              & Training         & 1350                   & 27                      \\
        1342227094\_L2\_5 & 50                  & 10                & 6000              & Training         & 1550                   & 31                      \\
        1342196757\_L2\_5 & 50                  & 10                & 6000              & Training         & 1550                   & 31                      \\
        1342226731\_L2\_5 & 50                  & 40                & 8000              & Training         & 1550                   & 31                      \\
        1342231674\_L2\_5 & 50                  & 40                & 6000              & Training         & 1450                   & 29                      \\
        1342205228\_L3    & 50                  & 900               & 40000             & Training         & 1650                   & 33                      \\
        1342228913\_L2\_5 & 50                  & 40                & 10000             & Training         & 1650                   & 33                      \\
        1342227049\_L2\_5 & 50                  & 100               & 11000             & Training         & 1550                   & 31                      \\
        1342218780\_L2\_5 & 30                  & 10                & 3000              & Training         & 780                    & 26                      \\
        \hline
        1342188204\_L2\_5 & 145                 & 20                & 6000              & Validation       & 4350                   & 30                      \\
        1342244881\_L2\_5 & 70                  & 20                & 7000              & Validation       & 2170                   & 31                      \\
        1342192767\_L3    & 70                  & 3500              & 40000             & Validation       & 1750                   & 25                      \\
        1342191819\_L2\_5 & 55                  & 400               & 35000             & Validation       & 1980                   & 36                      \\
        1342217448\_L2\_5 & 55                  & 150               & 14000             & Validation       & 1760                   & 32                      \\
        1342216507\_L2\_5 & 50                  & 10                & 3500              & Validation       & 1350                   & 27                      \\
        1342245206\_L3    & 45                  & 10                & 6000              & Validation       & 1395                   & 31                      \\
        \hline
        1342249954\_L2\_5 & 200                 & 100               & 12000             & Test             & 6400                   & 32                      \\
        1342212042\_L2\_5 & 135                 & 40                & 5500              & Test             & 3780                   & 28                      \\
        1342216409\_L2\_5 & 115                 & 40                & 3500              & Test             & 2875                   & 25                      \\
        1342237154\_L2\_5 & 90                  & 60                & 10000             & Test             & 2880                   & 32                      \\
        1342213538\_L3    & 40                  & 20                & 3000              & Test             & 1000                   & 25                      \\
        1342227047\_L2\_5 & 40                  & 1750              & 40000             & Test             & 1160                   & 29   \\
    \hline
    \end{tabular}
    \tablefoot{The columns from left to right represent: the observation ID (field); the number of sources per simulation injected into each simulation for this field; minimum and maximum flux density levels; the resulting dataset split (Training, Validation, or Test); the total number of sources (calculated as the number of simulations multiplied by the number of sources per simulation); and finally, the number of simulations, which corresponds to the range of flux levels between the minimum and maximum flux densities.}
\end{table*}

\begin{table*}[]
    \caption{Details of the dataset for the Scan Map Observing Mode in the Green band.}
    \label{table:scanmap_green_details}
    \centering
    \begin{tabular}{|c|c|c|c|c|c|c|}
    \hline
    \textbf{Field}         & \textbf{\# Sources} & \textbf{Min Flux} & \textbf{Max Flux} & \textbf{Dataset} & \textbf{Total Sources} & \textbf{\# Simulations} \\
    \hline
    1342188880\_L2\_5 & 190                 & 200               & 30000             & Training         & 7220                   & 38                      \\
    1342216587\_L2\_5 & 185                 & 40                & 7000              & Training         & 5550                   & 30                      \\
    1342227811\_L2\_5 & 185                 & 400               & 40000             & Training         & 7030                   & 38                      \\
    1342212040\_L2\_5 & 170                 & 20                & 12000             & Training         & 6120                   & 36                      \\
    1342204085\_L2\_5 & 170                 & 250               & 40000             & Training         & 6970                   & 41                      \\
    1342188882\_L2\_5 & 140                 & 150               & 30000             & Training         & 5600                   & 40                      \\
    1342185555\_L2\_5 & 75                  & 150               & 20000             & Training         & 2700                   & 36                      \\
    1342241313\_L3    & 65                  & 20                & 20000             & Training         & 2730                   & 42                      \\
    1342192771\_L3    & 60                  & 8000              & 40000             & Training         & 1140                   & 19                      \\
    1342187005\_L2\_5 & 60                  & 60                & 16000             & Training         & 2280                   & 38                      \\
    1342237430\_L2\_5 & 55                  & 40                & 18000             & Training         & 2200                   & 40                      \\
    1342191821\_L2\_5 & 50                  & 900               & 40000             & Training         & 1650                   & 33                      \\
    1342191807\_L2\_5 & 50                  & 200               & 22500             & Training         & 1750                   & 35                      \\
    1342256983\_L2\_5 & 45                  & 40                & 10000             & Training         & 1485                   & 33                      \\
    1342228954\_L2\_5 & 45                  & 80                & 22500             & Training         & 1800                   & 40                      \\
    1342263882\_L2\_5 & 40                  & 60                & 14000             & Training         & 1440                   & 36                      \\
    1342203102\_L2\_5 & 40                  & 60                & 15000             & Training         & 1480                   & 37                      \\
    1342209035\_L2\_5 & 35                  & 200               & 27500             & Training         & 1295                   & 37                      \\
    1342207059\_L2\_5 & 35                  & 150               & 25000             & Training         & 1330                   & 38                      \\
    1342204298\_L2\_5 & 35                  & 150               & 27500             & Training         & 1365                   & 39                      \\
    1342193498\_L2\_5 & 30                  & 40                & 8000              & Training         & 930                    & 31                      \\
    1342269038\_L2\_5 & 30                  & 80                & 14000             & Training         & 1050                   & 35                      \\
    1342193548\_L2\_5 & 30                  & 200               & 35000             & Training         & 1200                   & 40                      \\
    1342193520\_L2\_5 & 30                  & 150               & 35000             & Training         & 1260                   & 42                      \\
    1342191815\_L2\_5 & 30                  & 300               & 30000             & Training         & 1080                   & 36                      \\
    1342270796\_L2\_5 & 30                  & 100               & 18000             & Training         & 1110                   & 37                      \\
    1342185545\_L2\_5 & 30                  & 150               & 20000             & Training         & 1080                   & 36                      \\
    1342259541\_L2\_5 & 25                  & 20                & 6000              & Training         & 750                    & 30                      \\
    1342231368\_L2\_5 & 25                  & 60                & 8000              & Training         & 750                    & 30                      \\
    \hline
    1342218701\_L2\_5 & 185                 & 60                & 8000              & Validation       & 5550                   & 30                      \\
    1342241507\_L2\_5 & 100                 & 10                & 13000             & Validation       & 3800                   & 38                      \\
    1342267202\_L3    & 55                  & 80                & 25000             & Validation       & 2255                   & 41                      \\
    1342261768\_L2\_5 & 45                  & 40                & 9000              & Validation       & 1440                   & 32                      \\
    1342231676\_L2\_5 & 45                  & 60                & 20000             & Validation       & 1800                   & 40                      \\
    1342193494\_L2\_5 & 30                  & 150               & 27500             & Validation       & 1170                   & 39                      \\
    1342230107\_L2\_5 & 30                  & 150               & 20000             & Validation       & 1080                   & 36                      \\
    \hline
    1342237204\_L2\_5 & 135                 & 40                & 8000              & Test             & 4185                   & 31                      \\
    1342237206\_L2\_5 & 100                 & 200               & 40000             & Test             & 4200                   & 42                      \\
    1342188205\_L2\_5 & 100                 & 150               & 10000             & Test             & 2800                   & 28                      \\
    1342244224\_L2\_5 & 70                  & 40                & 7000              & Test             & 2100                   & 30                      \\
    1342189388\_L2\_5 & 40                  & 80                & 14000             & Test             & 1400                   & 35                      \\
    1342217764\_L2\_5 & 40                  & 250               & 25000             & Test             & 1400                   & 35  \\
    \hline
    \end{tabular}
    \tablefoot{The columns are organized as detailed in Table~\ref{table:scanmap_blue_details}.}
\end{table*}

\begin{table*}[]
    \caption{Details of the dataset for the Scan Map Observing Mode in the Red band.}
    \label{table:scanmap_red_details}
    \centering
    \begin{tabular}{|c|c|c|c|c|c|c|}
    \hline
    \textbf{Field}    & \textbf{\# Sources} & \textbf{Min Flux} & \textbf{Max Flux} & \textbf{Dataset} & \textbf{Total Sources} & \textbf{\# Simulations} \\
    \hline
    1342247320\_L3    & 180                 & 80                & 10000             & Training         & 5580                   & 31                      \\
    1342228445\_L2\_5 & 150                 & 150               & 15000             & Training         & 4950                   & 33                      \\
    1342215726\_L3    & 145                 & 1750              & 40000             & Training         & 4205                   & 29                      \\
    1342217454\_L2\_5 & 140                 & 80                & 8000              & Training         & 4060                   & 29                      \\
    1342221286\_L2\_5 & 140                 & 200               & 20000             & Training         & 4760                   & 34                      \\
    1342215595\_L2\_5 & 110                 & 150               & 16000             & Training         & 3740                   & 34                      \\
    1342220089\_L2\_5 & 90                  & 150               & 18000             & Training         & 3150                   & 35                      \\
    1342188204\_L3    & 85                  & 80                & 16000             & Training         & 3145                   & 37                      \\
    1342258816\_L2\_5 & 85                  & 500               & 40000             & Training         & 3145                   & 37                      \\
    1342188882\_L2\_5 & 85                  & 400               & 40000             & Training         & 3230                   & 38                      \\
    1342204085\_L2\_5 & 85                  & 700               & 40000             & Training         & 2975                   & 35                      \\
    1342195485\_L2\_5 & 85                  & 60                & 15000             & Training         & 3145                   & 37                      \\
    1342212040\_L3    & 85                  & 60                & 9000              & Training         & 2635                   & 31                      \\
    1342237206\_L2\_5 & 80                  & 300               & 40000             & Training         & 3200                   & 40                      \\
    1342224999\_L2\_5 & 65                  & 200               & 15000             & Training         & 2015                   & 31                      \\
    1342220308\_L2\_5 & 50                  & 250               & 15000             & Training         & 1500                   & 30                      \\
    1342220778\_L2\_5 & 50                  & 150               & 12000             & Training         & 1500                   & 30                      \\
    1342216401\_L3    & 45                  & 60                & 8000              & Training         & 1350                   & 30                      \\
    1342216501\_L2\_5 & 40                  & 80                & 10000             & Training         & 1240                   & 31                      \\
    1342222157\_L2\_5 & 40                  & 80                & 10000             & Training         & 1240                   & 31                      \\
    1342216415\_L2\_5 & 40                  & 80                & 10000             & Training         & 1240                   & 31                      \\
    1342241507\_L2\_5 & 40                  & 80                & 15000             & Training         & 1440                   & 36                      \\
    1342220784\_L2\_5 & 35                  & 100               & 7000              & Training         & 945                    & 27                      \\
    \hline
    1342249134\_L2\_5 & 165                 & 600               & 40000             & Validation       & 5940                   & 36                      \\
    1342220818\_L2\_5 & 90                  & 150               & 15000             & Validation       & 2970                   & 33                      \\
    1342219078\_L2\_5 & 90                  & 150               & 25000             & Validation       & 3420                   & 38                      \\
    1342218701\_L2\_5 & 60                  & 100               & 15000             & Validation       & 2100                   & 35                      \\
    \hline
    1342249269\_L3    & 130                 & 300               & 40000             & Test             & 5200                   & 40                      \\
    1342252015\_L2\_5 & 100                 & 125               & 14000             & Test             & 3300                   & 33                      \\
    1342223340\_L2\_5 & 90                  & 150               & 16000             & Test             & 3060                   & 34                      \\
    1342249136\_L3    & 70                  & 80                & 11000             & Test             & 2240                   & 32                      \\
    1342204168\_L2\_5 & 50                  & 80                & 7000              & Test             & 1400                   & 28 \\
    \hline
    \end{tabular}
    \tablefoot{The columns are organized as detailed in Table~\ref{table:scanmap_blue_details}.}
\end{table*}

\begin{table*}[]
\caption{Details of the dataset for the Parallel Observing Mode across all bands.}
\label{table:parallel_details}
\centering
\begin{tabular}{|c|c|c|c|c|c|c|c|}
\hline
\textbf{Band} & \textbf{Field}    & \textbf{\# Sources} & \textbf{Min Flux} & \textbf{Max Flux} & \textbf{Dataset} & \textbf{Total Sources} & \textbf{\# Simulations} \\
\hline
Blue          & 1342213208\_L2\_5 & 175                 & 200               & 30000             & Training         & 6650                   & 38                      \\
Blue          & 1342205093\_L2\_5 & 155                 & 600               & 40000             & Training         & 5580                   & 36                      \\
Blue          & 1342214504\_L2\_5 & 130                 & 200               & 40000             & Training         & 5460                   & 42                      \\
Blue          & 1342206679\_L2\_5 & 120                 & 200               & 30000             & Training         & 4560                   & 38                      \\
Blue          & 1342244188\_L2\_5 & 120                 & 200               & 35000             & Training         & 4800                   & 40                      \\
Blue          & 1342251926\_L2\_5 & 115                 & 200               & 30000             & Training         & 4370                   & 38                      \\
Blue          & 1342239278\_L2\_5 & 115                 & 100               & 30000             & Training         & 4830                   & 42                      \\
Blue          & 1342255011\_L2\_5 & 115                 & 500               & 40000             & Training         & 4255                   & 37                      \\
Blue          & 1342244168\_L2\_5 & 100                 & 2000              & 40000             & Training         & 2800                   & 28                      \\
Blue          & 1342255006\_L2\_5 & 90                  & 800               & 40000             & Training         & 3060                   & 34                      \\
Blue          & 1342189109\_L2\_5 & 90                  & 500               & 40000             & Training         & 3330                   & 37                      \\
Blue          & 1342244166\_L2\_5 & 80                  & 600               & 40000             & Training         & 2880                   & 36                      \\
Blue          & 1342204366\_L2\_5 & 80                  & 800               & 40000             & Training         & 2720                   & 34                      \\
Blue          & 1342255009\_L2\_5 & 75                  & 1000              & 40000             & Training         & 2400                   & 32                      \\
Blue          & 1342214761\_L2\_5 & 75                  & 600               & 40000             & Training         & 2700                   & 36                      \\
Blue          & 1342204368\_L2\_5 & 25                  & 400               & 40000             & Training         & 950                    & 38                      \\
\hline
Blue          & 1342244831\_L2\_5 & 125                 & 200               & 40000             & Validation       & 5250                   & 42                      \\
Blue          & 1342205078\_L2\_5 & 120                 & 200               & 30000             & Validation       & 4560                   & 38                      \\
Blue          & 1342188088\_L2\_5 & 90                  & 60                & 15000             & Validation       & 3330                   & 37                      \\
Blue          & 1342214510\_L2\_5 & 55                  & 400               & 40000             & Validation       & 2090                   & 38                      \\
\hline
Blue          & 1342185644\_L2\_5 & 60                  & 80                & 14000             & Test             & 2100                   & 35                      \\
Blue          & 1342257384\_L2\_5 & 85                  & 250               & 40000             & Test             & 3485                   & 41                      \\
Blue          & 1342257382\_L2\_5 & 115                 & 1000              & 40000             & Test             & 3680                   & 32                      \\
Blue          & 1342244170\_L2\_5 & 100                 & 600               & 40000             & Test             & 3600                   & 36                      \\
Blue          & 1342186121\_L2\_5 & 75                  & 350               & 20000             & Test             & 2325                   & 31                      \\
Blue          & 1342184484\_L2\_5 & 80                  & 600               & 40000             & Test             & 2880                   & 36                      \\
\hline
Green         & 1342189661\_L2\_5 & 210                 & 150               & 25000             & Training         & 7980                   & 38                      \\
Green         & 1342205047\_L2\_5 & 180                 & 200               & 30000             & Training         & 6840                   & 38                      \\
Green         & 1342187331\_L2\_5 & 170                 & 200               & 40000             & Training         & 7140                   & 42                      \\
Green         & 1342189845\_L2\_5 & 150                 & 200               & 30000             & Training         & 5700                   & 38                      \\
Green         & 1342188096\_L2\_5 & 140                 & 150               & 25000             & Training         & 5320                   & 38                      \\
Green         & 1342223213\_L3    & 60                  & 40                & 12000             & Training         & 2100                   & 35                      \\
\hline
Green         & 1342187170\_L2\_5 & 115                 & 250               & 25000             & Validation       & 4025                   & 35                      \\
Green         & 1342190275\_L2\_5 & 70                  & 80                & 18000             & Validation       & 2660                   & 38                      \\
Green         & 1342198592\_L2\_5 & 70                  & 100               & 25000             & Validation       & 2800                   & 40                      \\
\hline
Green         & 1342198246\_L2\_5 & 115                 & 150               & 30000             & Test             & 4600                   & 40                      \\
Green         & 1342197280\_L2\_5 & 75                  & 150               & 20000             & Test             & 2700                   & 36                      \\
Green         & 1342190273\_L2\_5 & 70                  & 80                & 25000             & Test             & 2870                   & 41                      \\
Green         & 1342239263\_L2\_5 & 65                  & 100               & 20000             & Test             & 2470                   & 38                      \\
Red           & 1342257382\_L2\_5 & 90                  & 600               & 40000             & Test             & 3240                   & 36                      \\
Red           & 1342188088\_L2\_5 & 80                  & 300               & 15000             & Test             & 2320                   & 29                      \\
Red           & 1342257384\_L2\_5 & 75                  & 300               & 40000             & Test             & 3000                   & 40                      \\
Red           & 1342186121\_L2\_5 & 70                  & 400               & 20000             & Test             & 2100                   & 30                      \\
Red           & 1342184484\_L2\_5 & 65                  & 600               & 40000             & Test             & 2340                   & 36                      \\
Red           & 1342185644\_L2\_5 & 55                  & 125               & 14000             & Test             & 1815                   & 33                     \\
\hline
\end{tabular}
\tablefoot{The columns from left to right represent: the band, observation ID (field), number of sources per simulation injected into each simulation of this field, minimum and maximum flux density levels, dataset split (Training, Validation, or Test), total number of sources (calculated as the number of simulations multiplied by the number of sources per simulation), and finally, the number of simulations, which corresponds to the range of flux levels between the minimum and maximum flux densities.}
\end{table*}
    
\end{appendix}

\end{document}